\documentclass[1p,number]{elsarticle}

\usepackage{sublabel}
\usepackage[]{graphicx}
\usepackage{epsf}
\usepackage{wrapfig}
\usepackage{epsfig}
\usepackage[usenames]{color}
\usepackage{pstricks}
\usepackage{amssymb, graphics}
\usepackage{indentfirst}
\usepackage{fancyhdr}
\usepackage{slashed}
\usepackage[latin1]{inputenc}
\usepackage[english]{babel}
\usepackage{amssymb}  
\usepackage{amsmath}  
\usepackage{latexsym} 
\usepackage{amsthm}   
\usepackage{axodraw4j}
\usepackage{subfigure}
\usepackage{color}

\journal{Nuclear Physics A}

\begin{document}
\title{Weak response of cold symmetric nuclear matter at three-body cluster level}
\author[sissa,infn]{Alessandro Lovato \corref{cor1}}
\ead{lovato@sissa.it}
\author[sissa]{Cristina Losa}
\author[sapienza,infn]{Omar Benhar}

\address[sissa]{ SISSA, I-34014 Trieste, Italy }
\address[infn]{INFN, Sezione di Roma. I-00185 Roma, Italy}
\address[sapienza]{Dipartimento di Fisica, Universit\`a ``La Sapienza''. I-00185 Roma, Italy}

\cortext[cor1]{Corresponding author}

\date{\today}
\begin{abstract}
We studied the Fermi and Gamow-Teller responses of cold symmetric nuclear matter within a unified dynamical model, suitable to account for both short- and long-range correlation effects. The formalism of correlated basis functions has been used to construct two-body effective interactions and one-body effective weak operators. The inclusion of the three-body cluster term allowed for incorporating in the effective interaction a realistic model of three- nucleon forces, namely the UIX potential. Moreover, the sizable unphysical dependence of the effective weak operator is removed once the three-body cluster term is taken into account. 
\end{abstract}


\maketitle


\section{Introduction}

The understanding of the interaction of low-energy neutrinos with nuclear matter is required for a description of a number of properties of compact stars.

The neutrino mean free path is thought to play a crucial role in the mechanism leading to supernovae explosion, while neutrino emission is the main process driving the early stages of neutron stars' cooling. 

The charged-current weak response of isospin symmetric nuclear matter has been analyzed within the framework of correlated basis functions (CBF) theory in Refs. \cite{cowell_04,benhar_09}. Unlike the earlier applications of the CBF formalism to the density and electromagnetic weak response \cite{fantoni_87,fabrocini_89,fabrocini_97}, the works of Refs. \cite{cowell_04,benhar_09} are based on effective interactions and effective transition operators, allowing for a consistent treatment of short-range correlations, which are known to be dominant at large momentum transfer, and long-range correlations, leading to the excitations of collective modes at low momentum transfer. 

The effective interaction of Ref. \cite{cowell_04} has been obtained at two-body cluster level, in the cluster expansion of the ground-state expectation value of an hamiltonian including a truncated version of the Argonne $v_{18}$ potential \cite{wiringa_95}. The authors of Ref. \cite{benhar_09} also took into account the effects of interactions involving three or more nucleons through the density dependent phenomenological potential originally developed in Ref. \cite{lagaris_81}. In this work we improve the CBF effective interaction by explicitly including the three-body cluster contributions, which allows for a more realistic description of three-nucleon forces at microscopic level. In particular we have been able to include the leading contributions of the UIX three-body interaction \cite{pudliner_95}. The effective transition operators are also consistently evaluated including the leading order three-body cluster terms. 

An overview of the formalism is given in Section \ref{sec:formalism}.  In Section \ref{sec:ei} the effective interaction at three-body cluster level for an hamiltonian with Argonne $v_{6}^\prime\,+\,UIX$ potential is derived. Section \ref{sec:ewo} is devoted to the development of the weak transition operator at three-body cluster level. The correlated Fermi gas (CFG), the correlated Hartree-Fock (CHF), and the correlated Tamm-Dancoff (CTD) approximations are discussed in Section \ref{sec:resp_appr}. The numerical calculations of the response and of its sum rule are reported in Section \ref{sec:num_calc}. Finally, conclusions and future perspectives are discussed in Section \ref{sec:concl}.

\section{Formalism}
\label{sec:formalism}
In the low-momentum transfer regime ($|\mathbf{q}|$ of the order of 10 MeV), the non relativistic limit of the weak-charged current matrix element is expected to be applicable. The nuclear response to weak probes delivering energy $\omega$ and momentum \textbf{q} at leading order in $|\mathbf{q}|/m$ reads
\begin{align}
S(\textbf{q},\omega)=\frac{1}{A}\sum_{n\neq0} |\langle \Psi_n| \hat{O}_{\textbf{q}}| \Psi_0\rangle |^2 \delta(\omega+E_0-E_n)\,.
\label{eq:resp_def}
\end{align}
In the above equation $A$ is the particle number and $\hat{O}_{\textbf{q}}$ the one--body weak operator that induces a transition from the ground state $|\Psi_0\rangle$ to the excited state $|\Psi_n\rangle$, which are eigenstates of the nuclear hamiltonian with energies $E_0$ and $E_n$, respectively, i.e.
\begin{equation}
\hat{H}|\Psi_n\rangle=E_n |\Psi_n\rangle\, .
\end{equation}

The non relativistic Fermi (F) and Gamow-Teller (GT) operators describing low energy weak interactions are
\begin{align}
\hat{O}_{\mathbf{q}}^F &= \sum_i \hat{O}_{\mathbf{q}}^F(i)=g_V\sum_i e^{i \mathbf{q} \cdot \mathbf{r}_i}\tau^{+}_i,\\
\hat{O}_{\mathbf{q}}^{GT} &= \sum_i \hat{O}_{\mathbf{q}}^{GT}(i)=g_A\sum_i e^{i \mathbf{q} \cdot \mathbf{r}_i} \vec{\sigma}_i \tau^{+}_i,
\label{eq:f_gt_def}
\end{align}
where $g_V=1.00$ and $g_A=1.26$ are the form factors at zero momentum transfer, while $\tau^{+}_i$ is the isospin-raising operator acting on the $i$-th
nucleon. 

The spin-longitudinal and spin-transverse components of the Gamow-Teller response functions are defined as the components parallel and orthogonal to $\mathbf{q}$, respectively. They can differ significantly at large values of $|{\bf q}|$ and, in principle, should be calculated separately. However, whenever not otherwise specified, with ``Gamow-Teller response'' we will be referring to the total response, defined as the sum of the cartesian components
\begin{align}
S^{GT}(\textbf{q},\omega)=&\frac{1}{A}\sum_{\alpha=x,y,z} \sum_n |\langle \Psi_n | \hat{O}_{\textbf{q}\,\alpha}^{GT}| \Psi_0\rangle |^2  \delta(\omega+E_0-E_n)\,.
\label{eq:resp_gttot}
\end{align}

The differences between the integrated spin-longitudinal and spin-transverse responses will be discussed in Section \ref{sec:sum_rule}.

Within CBF, the states appearing in Eq. (\ref{eq:resp_def}) are written in the form \cite{clark_66,fantoni_98}
\begin{equation}
|\Psi_n\rangle \equiv\frac{\hat{\mathcal{F}}|\Phi_n\rangle}{\langle \Phi_n|\hat{\mathcal{F}}^\dagger \hat{\mathcal{F}} | \Phi_n\rangle},
\end{equation}
where $|\Phi_n\rangle$ is the Slater determinant of non interacting $n-$particle $n-$hole state. The structure of the correlation operator, $\hat{\mathcal{F}}$, reflects the complexity of the Argonne $v_{6}^\prime$ nucleon-nucleon potential \cite{pandharipande_71,ristig_71,pandharipande_72}
\begin{equation}
\mathcal{F}=\mathcal{S} \prod_{j>i=1}^A \hat{F}_{ij} \ ,
\label{eq:Foperator}
\end{equation}
with
\begin{align}
\hat{F}_{ij}& = \sum_{p=1}^6 f^{p}(r_{ij})\hat{O}^{p}_{ij}
\label{eq:Foperator_v6prime}
\end{align}
and
\begin{align}
\hat{O}^{p=1-6}_{ij}&=(1,\sigma_{ij},S_{ij})\otimes(1,\tau_{ij})\,.
\end{align}
In the above equation,  $\sigma_{ij}=\vec{\sigma}_i \cdot \vec{\sigma}_j$ and $\tau_{ij}=\vec{\tau}_i \cdot \vec{\tau}_j$, where $\vec{\sigma}_i$ and $\vec{\tau}_i$ are Pauli matrices acting on the spin or isospin of the $i$-th, while
\begin{equation}
S_{ij}=(3\hat{r}_{ij}^\alpha\hat{r}_{ij}^\beta-\delta^{\alpha\beta})\sigma_{i}^\alpha\sigma_{j}^\beta \ ,
\label{eq:tens_def}
\end{equation}
with $\alpha, \ \beta= 1, \ 2, \ 3$, is the tensor operator.

Note that the symmetrization operator $\mathcal{S}$ is needed to fulfill the requirement of antisymmetrization of the state 
$|\Psi_n\rangle$, since, in general, $[\hat{O}^{p}_{ij},\hat{O}^{q}_{ik}] \neq 0$. 

The variational parameters determining the shape of the radial functions $f^{p}(r_{ij})$ have been fixed in Ref. \cite{lovato_11}, minimizing the variational ground-state energy. In this work we will use the results corresponding to the hamiltonians with Argonne $v_{6}^\prime$ and Argonne $v_{6}^\prime$ + UIX potentials.

Following Refs. \cite{cowell_04,benhar_09}, we only consider transitions between the correlated ground-state and correlated 1particle-1hole ($1p-1h$) excited states. The $np-nh$ states with $n\geq 2$ give a smaller contribution, mainly at large excitation energy.
the weak response. 

The CBF matrix element between the ground-state and $1p-1h$ excitation reads
\begin{align}
&\langle \Psi_{p_m;h_i}|\hat{O}_\mathbf{q}|0\rangle=\frac{\langle \Phi_{p_m; h_i}| \mathcal{F}^\dagger \hat{O}_\mathbf{q}\mathcal{F} | \Phi_0\rangle}{\sqrt{\langle \Phi_0 | \mathcal{F}^\dagger \mathcal{F} | \Phi_0\rangle\langle \Phi_{p_m; h_i}| \mathcal{F}^\dagger \mathcal{F} | \Phi_{p_m; h_i}\rangle}}\, ,
\label{eq:cbf_1p1h}
\end{align}
where $p_m$ and $h_i$ denote the whole set of quantum numbers of the single nucleon state, namely the momentum, the spin and the isospin projections along the $z-$axis.

This quantity, entering all our calculations of the response function, will allow us to define the effective weak operators, as discussed in Section \ref{sec:ewo}.

\section{Effective interaction}
\label{sec:ei}
Using the formalism of CBF and the cluster expansion technique, the authors of Ref. \cite{cowell_03,cowell_04}, were able to develop an {\it effective interaction}, obtained from the bare Argonne $v_{8}^\prime$ potential, which incorporates the effects of the short-range correlations. In Ref. \cite{benhar_07}, the two-body effective interaction 
of Ref. \cite{cowell_03,cowell_04} was improved with the inclusion of the purely phenomenological density dependent potential of Ref. \cite{lagaris_80}, 
accounting for the effects of interactions involving more than two nucleons. The CBF effective interaction, $v_{12}^{eff}$, suitable for use in Hartree-Fock calculations, is defined through the matrix elements of the hamiltonian in the correlated ground-state
\begin{equation}
\langle\Psi_0|\hat{\mathcal{F}}^\dag \hat{H} \hat{\mathcal{F}}|\Psi_0\rangle\equiv T_F+\langle \Phi_0 | \hat{v}_{12}^{eff} | \Phi_0 \rangle\, .
\label{eq:eff_int}
\end{equation}

As suggested by the above equation, the effective interaction allows one to calculate any nuclear matter observables using perturbation theory in the orthonormal FG basis. However, in general, extracting the effective interaction is a very challenging task, involving difficulties even more severe than those associated with the calculation of the expectation value of the hamiltonian in the correlated ground state.

\subsection{Two-body cluster}

The procedure developed in Ref. \cite{cowell_04} consists in carrying out a cluster expansion of the {\it lhs} of Eq. (\ref{eq:eff_int})
and keeping only the two-body cluster contribution. The sum of the two-body cluster contribution of the potential and kinetic energies are readily found to be
\begin{align}
\langle\Phi_0|\hat{\mathcal{F}}^\dag \hat{H} \hat{\mathcal{F}}|\Phi_0\rangle\Big{|}_{2b}=&
\frac{\rho}{2}\int d\mathbf{r}_{12}\text{CTr}\Big[\Big(\hat{F}_{12}\hat{v}_{12}\hat{F}_{12}-\frac{\hbar^2}{m}(\vec{\nabla}_1\hat{F}_{12})(\vec{\nabla}_{1}\hat{F}_{12})\Big)\nonumber\\
&\times(1-\hat{P}^{\sigma\tau}_{12}\ell_{12}^2)\Big]\, ,
\label{eq:e2bc}
\end{align}
where the symbol ``CTr'' denoting the normalized trace over the spin-isospin degrees of freedom of particles $1$, $2$ and $3$
The spin-isospin exchange operator and the Slater function are defined as
\begin{align}
\hat{P}_{ij}^{\sigma\tau}&=\frac{1}{4}(1+\sigma_{ij})(1+\tau_{ij})\nonumber\\
\ell(r_{ij})&=3\Big[\frac{\sin(k_Fr_{ij})-k_Fr_{ij}\cos(k_Fr_{ij})}{(k_Fr_{ij})^3}\Big]\,,
\end{align}
being $k_F$ the Fermi momentum of the system.

On the other hand, the expectation value of the effective potential is given by
\begin{align}
&\langle\Phi_0|\hat{v}^{eff}_{12}|\Phi_0\rangle=\frac{\rho}{2}\int d\mathbf{r}_{12} \text{CTr}\Big[
\hat{v}^{eff}_{12}\Big(1-\hat{P}_{12}\ell_{12}^2)\Big]\, .
\label{eq:eff_pot}
\end{align}
Therefore, the effective potential at two-body cluster level turns out to be
\begin{equation}
\hat{v}_{12}^{eff}\Big|_{2b}=\hat{F}v_{12}\hat{F}-\frac{\hbar^2}{m}(\vec{\nabla}_1\hat{F}_{12})(\vec{\nabla}_{1}\hat{F}_{12})\, .\end{equation}

The effective potential of the above equation slightly differs from the one reported in the literature. The authors of Refs. \cite{cowell_04,benhar_09} have not done the integration by parts leading to the kinetic term of Eq. (\ref{eq:e2bc}). As a consequence they have neglected the terms in which the gradient operates on both the correlation function and on the plane waves. In our effective potential, these terms, although small compared to the other contributions, are fully taken into account.

\subsection{Three-body cluster}
We have improved the effective potential by adding the tree-body cluster contributions to the energy per particle firstly calculated in Ref. \cite{morales_02}. This allowed us to consistently include in the effective interaction the UIX potential, whose leading order terms emerge at three-body cluster level. 

The three-body cluster contribution appearing in the expansion of $\mathcal{F}^\dagger v_{12} \mathcal{F}$ is given by
\begin{align}
\mathcal{F}^\dagger v_{12} \mathcal{F}\,\Big{|}_{3b}=&\sum_{i>2}\Big[\Big(\mathcal{S}\hat{F}_{12}\hat{F}_{1i}\hat{F}_{2i}\Big)\hat{v}_{12}\Big(\mathcal{S}\hat{F}_{12}\hat{F}_{1i}\hat{F}_{2i}\Big)-\hat{F}_{12}\hat{v}_{12}\hat{F}_{12}\Big]\, .
\end{align}
Within the FR diagrammatic scheme, the three-body cluster contribution to $\langle \hat{v}_{12} \rangle$ it is not merely the expectation value of the latter result, unlike the two-body case. As a matter of fact, the reducible diagrams arising form four-body cluster term of  $\mathcal{F}^\dagger v_{12} \mathcal{F}$, the detailed calculations of which can be found in Ref. \cite{lovato_12c}, needs to be taken into account. 

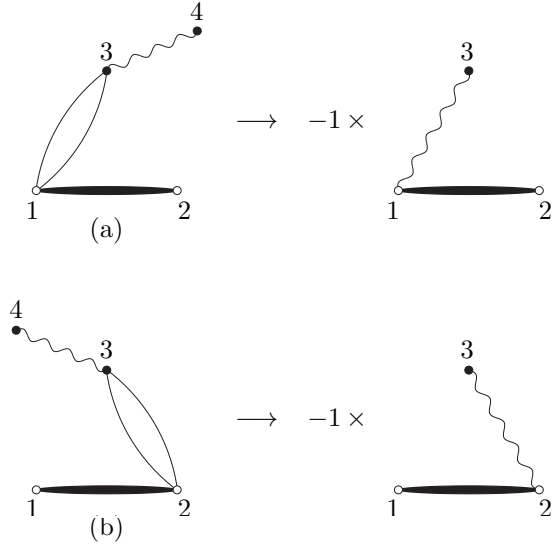
\begin{figure}[!h]
\begin{center}
\fcolorbox{white}{white}{
  \begin{picture}(200,200)(-10,-120)
	\SetWidth{0.5}
	\SetScale{0.75}	
        \unitlength=0.75 pt
	
	\CArc(90,-12.2917)(91,127,172)
        \CArc(-55,72.2917)(91,307,353)
	\COval(35,0)(2,32)(0){Black}{Black}
	\Photon(35,60)(80,80){2}{4}
        \BCirc(0,0){2}
	\CCirc(35,60){2}{Black}{Black}
	\BCirc(70,0){2}
        \CCirc(80,80){2}{Black}{Black}
	\Text(-2,-10)[]{1}
	\Text(74,-10)[]{2}
	\Text(35,70)[]{3}
	\Text(80,90)[]{4}
	\Text(35,-20)[]{(a)}

        \Text(110,34)[]{$\longrightarrow$}
        
        \Text(150,35)[]{$-1\,\times$}
	\Photon(180,0)(215,60){2}{5}
	\COval(215,0)(2,32)(0){Black}{Black}
        \BCirc(180,0){2}
	\CCirc(215,60){2}{Black}{Black}
	\BCirc(250,0){2}
	\Text(178,-10)[]{1}
	\Text(254,-10)[]{2}
	\Text(215,70)[]{3}
	
	\CArc(-20,-162.2917)(91,8,52)
        \CArc(125,-77.7083)(91,188,233)
	\COval(35,-150)(2,32)(0){Black}{Black}
	\Photon(35,-90)(-10,-70){2}{4}
        \BCirc(0,-150){2}
	\CCirc(35,-90){2}{Black}{Black}
	\BCirc(70,-150){2}
        \CCirc(-10,-70){2}{Black}{Black}

	\Text(-2,-160)[]{1}
	\Text(74,-160)[]{2}
        \Text(35,-80)[]{3}
	\Text(-10,-60)[]{4}
	\Text(35,-170)[]{(b)}

        \Text(110,-116)[]{$\longrightarrow$}
        \Text(150,-115)[]{$-1\,\times$}

	\Photon(250,-150)(215,-90){2}{5}
	\COval(215,-150)(2,32)(0){Black}{Black}
        \BCirc(180,-150){2}
	\CCirc(215,-90){2}{Black}{Black}
	\BCirc(250,-150){2}
	\Text(178,-160)[]{1}
	\Text(254,-160)[]{2}
	\Text(215,-80)[]{3}
	
\end{picture}
}
\vspace{1cm}
\caption{Four-body reducible diagrams, $v_{4b\to3b}^{\text{dir}}$ and their three-body reduction.\label{fig:red4bdir}}
\end{center}
\end{figure}

The direct term of the three-body cluster contribution in the FR expansion scheme is given by 
\begin{align}
\langle \hat{v}_{12} \rangle\Big{|}_{3b}^{\text{dir}}=&\frac{\rho^2}{2}\int d\mathbf{r}_{12}d\mathbf{r}_{13} \text{CTr}\Big[\nonumber \\
&(\mathcal{S}\hat{F}_{12}\hat{F}_{13}\hat{F}_{23})\hat{v}_{12}(\mathcal{S}\hat{F}_{12}\hat{F}_{13}\hat{F}_{23})-\hat{F}_{12}\hat{v}_{12}\hat{F}_{12}(\hat{F}_{13}^2+\hat{F}_{23}^2-1)\Big]\,.
\label{eq:v3b_dir}
\end{align}
It includes the term
\begin{equation}
(\mathcal{S}\hat{F}_{12}\hat{F}_{13}\hat{F}_{23})\hat{v}_{12}(\mathcal{S}\hat{F}_{12}\hat{F}_{13}\hat{F}_{23})-\hat{F}_{12}\hat{v}_{12}\hat{F}_{12}\, 
\end{equation}
from the three-body cluster contribution of $\mathcal{F}^\dagger v_{12} \mathcal{F}$, whereas the reducible four-body diagrams of Fig. \ref{fig:red4bdir} contribute with the factor 
\begin{equation}
-\hat{F}_{12}\hat{v}_{12}\hat{F}_{12}(\hat{F}_{13}^2+\hat{F}_{23}^2-2)\, .
\end{equation}

\begin{figure}[!ht]
\begin{center}
\fcolorbox{white}{white}{
  \begin{picture}(200,80)(-10,5)
	\SetWidth{0.5}
	\SetScale{0.75}	
        \unitlength=0.75 pt
        
	\CArc(90,-12.2917)(91,127,172)
	\CArc(-20,-12.2917)(91,7,52)
        \CArc(35,71)(80,245,295)
	\COval(35,0)(2,32)(0){Black}{Black}
	\Photon(35,60)(80,80){2}{4}
        \BCirc(0,0){2}
	\CCirc(35,60){2}{Black}{Black}
	\BCirc(70,0){2}
        \CCirc(80,80){2}{Black}{Black}
	\Text(-2,-10)[]{1}
	\Text(74,-10)[]{2}
	\Text(35,70)[]{3}
	\Text(80,90)[]{4}

        \Text(110,34)[]{$\longrightarrow$}
        
        \Text(150,35)[]{$-1\,\times$}
	\Photon(180,0)(215,60){2}{5}
	\CArc(215,71)(80,245,295)
        \CArc(215,-71)(80,65,115)

	\COval(215,0)(2,32)(0){Black}{Black}
        \BCirc(180,0){2}
	\CCirc(215,60){2}{Black}{Black}
	\BCirc(250,0){2}
	\Text(178,-10)[]{1}
	\Text(254,-10)[]{2}
	\Text(215,70)[]{3}
\end{picture}
}
\vspace{1.5cm}
\caption{Four-body reducible diagram, $v_{4b\to3b}^{\text{P12}}$, and its three-body reduction.\label{fig:red4bp12}}
\end{center}
\end{figure}

The sum of the diagrams where particles $1$ and $2$ are exchanged gives
\begin{align}
\langle \hat{v}_{12} \rangle\Big{|}_{3b}^{P_{12}}=&-\frac{\rho^2}{2}\int d\mathbf{r}_{12}d\mathbf{r}_{13}\ell^{2}_{12} \text{CTr}\Big[\nonumber\\
&(\mathcal{S}\hat{F}_{12}\hat{F}_{13}\hat{F}_{23})\hat{v}_{12}(\mathcal{S}\hat{F}_{12}\hat{F}_{13}\hat{F}_{23})\hat{P}^{\sigma\tau}_{12}-\hat{F}_{12}\hat{v}_{12}\hat{F}_{12}(\hat{F}_{13}^2+\hat{F}_{23}^2-1)\hat{P}^{\sigma\tau}_{12}\Big]\,.
\label{eq:v3b_p12}
\end{align}
The corresponding four-body diagram producing the term $\frac{1}{2}\hat{F}_{12}\hat{v}_{12}\hat{F}_{12}(\hat{F}_{13}^2+\hat{F}_{23}^2-2)\hat{P}^{\sigma\tau}_{12}$ is drawn in  Fig. \ref{fig:red4bp12}

\begin{figure}[!b]
\begin{center}
\fcolorbox{white}{white}{
  \begin{picture}(200,80)(-10,5)
	\SetWidth{0.5}
	\SetScale{0.75}	
        \unitlength=0.75 pt
	
	 \CArc(-55,72.2917)(91,307,353)
        \CArc(76,50.125)(91,160,214)
        \CArc(0,41.875)(40,30,105)
	\COval(35,0)(2,32)(0){Black}{Black}
	\Photon(35,60)(-10,80){2}{4}
        \BCirc(0,0){2}
	\CCirc(35,60){2}{Black}{Black}
	\BCirc(70,0){2}
        \CCirc(-10,80){2}{Black}{Black}
	\Text(-2,-10)[]{1}
	\Text(74,-10)[]{2}
	\Text(36,70)[]{3}
	\Text(-10,90)[]{4}

        \Text(110,34)[]{$\longrightarrow$}
        \Text(150,35)[]{$-1\,\times$}

	\Photon(180,0)(215,60){2}{5}
	\CArc(270,-12.2917)(91,127,172)
        \CArc(125,72.2917)(91,307,353)
	\COval(215,0)(2,32)(0){Black}{Black}
        \BCirc(180,0){2}
	\CCirc(215,60){2}{Black}{Black}
	\BCirc(250,0){2}
	\Text(178,-10)[]{1}
	\Text(254,-10)[]{2}
	\Text(215,70)[]{3}
\end{picture}
}
\vspace{1.5cm}
\caption{Four-body reducible diagram, $v_{4b\to3b}^{\text{P13}}$, and its three-body reduction.\label{fig:red4bp13}}
\end{center}
\end{figure}
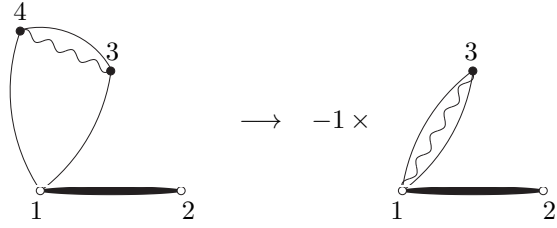

The diagrams in which particles $1$ and $3$ are exchanged contributes with
\begin{align}
\langle \hat{v}_{12} \rangle\Big{|}_{3b}^{P_{13}}=&-\frac{\rho^2}{2}\int d\mathbf{r}_{12}d\mathbf{r}_{13}\ell^{2}_{13} \text{CTr}\Big[\nonumber\\
&(\mathcal{S}\hat{F}_{12}\hat{F}_{13}\hat{F}_{23})\hat{v}_{12}(\mathcal{S}\hat{F}_{12}\hat{F}_{13}\hat{F}_{23})\hat{P}^{\sigma\tau}_{13}-\hat{F}_{12}\hat{v}_{12}\hat{F}_{12}\hat{F}_{13}^2\hat{P}^{\sigma\tau}_{13}\Big]\,.
\label{eq:v3b_p13}
\end{align}
where the term $\frac{1}{2}\hat{F}_{12}\hat{v}_{12}\hat{F}_{12}(\hat{F}_{13}^2-1)\hat{P}^{\sigma\tau}_{13}$ comes from the four-body reducible diagram of Fig. \ref{fig:red4bp13}.

Since the potential is invariant under $x_1\leftrightarrow x_2$. the diagrams with the exchange between particles $2$ and $3$ give the same contribution reported in Eq. (\ref{eq:v3b_p13}). The associated four-body reducible diagram is very similar to the one of Fig. \ref{fig:red4bp13} but with the loop attached to particle $2$ instead of particle $1$.

Consider the diagrams with the circular exchange involving particles $1$, $2$ and $3$. In this case there are no reducible four-body diagrams that partly cancel the reducible part of the three body diagram. In addition, there are no three-body reducible diagrams with circular exchange at all. However, the four-body diagram of Fig. \ref{fig:red4bp13}, with no correlation lines linking particles $1$ and $2$ to the others, can be {\it reduced} to a three-body term, so that the three-body diagram with a circular exchange reads
\begin{align}
\langle \hat{v}_{12} \rangle\Big{|}_{3b}^{cir}=&\rho^2\int d\mathbf{r}_{12}d\mathbf{r}_{13}\ell_{12}\ell_{13}\ell_{23} \text{CTr}\Big[\nonumber \\
&(\mathcal{S}\hat{F}_{12}\hat{F}_{13}\hat{F}_{23})\hat{v}_{12}(\mathcal{S}\hat{F}_{12}\hat{F}_{13}\hat{F}_{23})\hat{P}^{\sigma\tau}_{12}\hat{P}^{\sigma\tau}_{13}-\hat{F}_{12}\hat{v}_{12}\hat{F}_{12}\hat{F}_{13}^2\hat{P}^{\sigma\tau}_{13}\hat{P}^{\sigma\tau}_{12}\Big]\,.
\label{eq:v3b_cir}
\end{align}

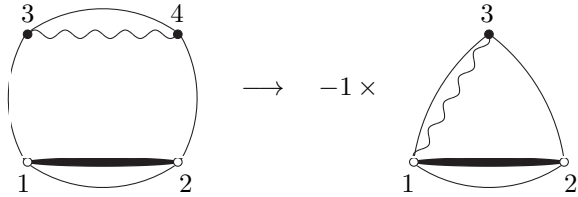
\begin{figure}[!hb]
\begin{center}
\fcolorbox{white}{white}{
  \begin{picture}(200,80)(0,5)
	\SetWidth{0.5}
	\SetScale{0.8}	
        \unitlength=0.8 pt
	
	\CArc(51,30)(60,148,210)
        \CArc(19,30)(60,-30,32)
        \CArc(35,48)(60,236,304)
        \CArc(35,12)(60,56,124)
	\COval(35,0)(2,32)(0){Black}{Black}
	\Photon(0,60)(70,60){2}{5}
        \BCirc(0,0){2}
	\CCirc(0,60){2}{Black}{Black}
	\BCirc(70,0){2}
        \CCirc(70,60){2}{Black}{Black}
	\Text(-2,-10)[]{1}
	\Text(74,-10)[]{2}
	\Text(0,70)[]{3}
	\Text(70,70)[]{4}

        \Text(110,34)[]{$\longrightarrow$}
        \Text(150,35)[]{$-1\,\times$}

	\Photon(180,0)(215,60){2}{5}
	\CArc(270,-12.2917)(91,127,172)
        \CArc(160,-12.2917)(91,7,53)
        \CArc(215,48)(60,236,304)

	\COval(215,0)(2,32)(0){Black}{Black}
        \BCirc(180,0){2}
	\CCirc(215,60){2}{Black}{Black}
	\BCirc(250,0){2}
	\Text(178,-10)[]{1}
	\Text(254,-10)[]{2}
	\Text(215,70)[]{3}
\end{picture}
}
\vspace{1.5cm}
\caption{Four-body diagram, $v_{4b\to3b}^{\text{cir}}$, that contributes to the three-body diagrams having a circular exchange between particles $1$, $2$ and $3$. \label{fig:red4bcir}}
\end{center}
\end{figure}

The three-body cluster contribution to the PB kinetic energy contains terms of the kind $\nabla_{1}^2(\mathcal{S} \hat{F}_{12}\hat{F}_{13}\hat{F}_{23})$. Their explicit expressions can be obtained from the corresponding equations for the two-body potential by substituting the first term of the normalized traces with \cite{morales_02}
\begin{align}
\hat{v}_{12}(\mathcal{S} \hat{F}_{12}\hat{F}_{13}\hat{F}_{23})\to&-2[\mathcal{S} (\nabla_{1}^2\hat{F}_{12})\hat{F}_{13}\hat{F}_{23}]-2[\mathcal{S} (\vec{\nabla}_{1}\hat{F}_{12})\cdot(\vec{\nabla}_1\hat{F}_{13})\hat{F}_{23}]\, ,
\end{align}
while
\begin{equation}
\hat{v}_{12} \hat{F}_{12}\to-2 (\nabla_{1}^2\hat{F}_{12})
\end{equation}
for the second term. Terms with $(\nabla_{1}^2\hat{F}_{12})$ are denoted by $W^{kin}$, those having $(\vec{\nabla}_{1}\hat{F}_{12})\cdot(\vec{\nabla}_1\hat{F}_{13})$ are included in $U$. On the other hand, the three-body cluster terms belonging to $W_F$ arise from the diagrams where particles $1$ and $2$ are exchanged 
\begin{align}
\langle \hat{T} \rangle_{W_F}\Big{|}_{3b}^{P_{12}}=&\frac{\hbar^2}{m}\rho^2\int d\mathbf{r}_{12}d\mathbf{r}_{13}\ell_{12}\ell^{\prime}_{12}\hat{r}_{12}\cdot \text{CTr}\Big[\nonumber\\
&(\mathcal{S}\hat{F}_{12}\hat{F}_{13}\hat{F}_{23})[\mathcal{S}(\vec{\nabla}_1\hat{F}_{12})\hat{F}_{13}\hat{F}_{23})\hat{P}^{\sigma\tau}_{12}-\hat{F}_{12}(\vec{\nabla}_{1}\hat{F}_{12})(\hat{F}_{13}^2+\hat{F}_{23}^2-1)\hat{P}^{\sigma\tau}_{12}\Big]
\label{eq:wf_p12}
\end{align}
and from the ones with circular exchange
\begin{align}
\langle \hat{T}\rangle_{W_F}\Big{|}_{3b}^{cir}=&-\frac{\hbar^2}{m}\rho^2\int d\mathbf{r}_{12}d\mathbf{r}_{13}\ell_{13}\ell_{23}\ell^{\prime}_{12}\hat{r}_{12}\cdot \text{CTr}\Big[\nonumber \\
&(\mathcal{S}\hat{F}_{12}\hat{F}_{13}\hat{F}_{23})[\mathcal{S}(\vec{\nabla}_1\hat{F}_{12})\hat{F}_{13}\hat{F}_{23}]\hat{P}^{\sigma\tau}_{12}\hat{P}^{\sigma\tau}_{13}-\hat{F}_{12}(\vec{\nabla}_{1}\hat{F}_{12})\hat{F}_{13}^2\hat{P}^{\sigma\tau}_{13}\hat{P}^{\sigma\tau}_{12}\Big]\,.
\label{eq:wf_cir}
\end{align}

The contributions to $U$ stem from the diagrams with the exchange $P_{13}$
\begin{align}
\langle \hat{T} \rangle_{U_F}\Big{|}_{3b}^{P_{13}}=&\frac{\hbar^2}{m}\rho^2\int d\mathbf{r}_{12}d\mathbf{r}_{13}\ell '(r_{13})\hat{r}_{13}\cdot \text{CTr}\Big[\nonumber\\
&(\mathcal{S}\hat{F}_{12}\hat{F}_{13}\hat{F}_{23})[\mathcal{S}(\vec{\nabla}_1\hat{F}_{12})\hat{F}_{13}\hat{F}_{23})\hat{P}^{\sigma\tau}_{13}\Big]
\label{eq:uf_p12}
\end{align}
and from those having circular exchange
\begin{align}
\langle \hat{T} \rangle_{U_F}\Big{|}_{3b}^{cir}=&-\frac{\hbar^2}{m}\rho^2\int d\mathbf{r}_{12}d\mathbf{r}_{13}\ell_{12}\ell_{13}\ell ^{\prime}_{13}\hat{r}_{23}\cdot\text{CTr}\Big[\nonumber\\
&(\mathcal{S}\hat{F}_{12}\hat{F}_{13}\hat{F}_{23})[\mathcal{S}(\vec{\nabla}_1\hat{F}_{12})\hat{F}_{13}\hat{F}_{23}]\hat{P}^{\sigma\tau}_{12}\hat{P}^{\sigma\tau}_{13}\Big]\,.
\label{eq:uf_cir}
\end{align}
Note that in this case there are no subtraction terms arising from reducible diagrams.

Finally, the three-body cluster term of the three-body potential $\hat{V}_{123}$, like UIX model, can be easily obtained by keeping
only the first term in the traces of Eqs. (\ref{eq:v3b_dir}), (\ref{eq:v3b_p12}), (\ref{eq:v3b_p13}) and (\ref{eq:v3b_cir}) with the following replacement
\begin{equation}
\hat{v}_{12}\to \frac{\hat{V}_{123}}{3}\, .
\end{equation}

As in the construction of the density dependent potential from UIX, developed in Ref. \cite{lovato_11}, the issue of the exchange pattern has to be carefully analyzed. The distinctive feature of the present calculation is that $v_{12}^{eff}$ contains the correlation between particles $1$ and $2$ making possible to implement the inversion of $\hat{P}^{\sigma\tau}_{ij}$ in a straightforward way. To be definite, consider the three-body cluster contribution of the ground-state expectation value of the two-body potential
\begin{align}
\langle \hat{v}_{12} \rangle\Big{|}_{3b}=&\frac{\rho^2}{2}\int dx_{123}\hat{X}(x_1,x_2;x_3)\Big[1-\hat{P}^{\sigma\tau}_{12}\ell^{2}_{12}-2\hat{P}^{\sigma\tau}_{13}\ell^{2}_{13}+2\hat{P}^{\sigma\tau}_{12}\hat{P}^{\sigma\tau}_{13}\ell_{12}\ell_{13}\ell_{23})\Big]\, .
\end{align}

\begin{figure}[!ht]
\begin{center}
\includegraphics[width=8.0cm]{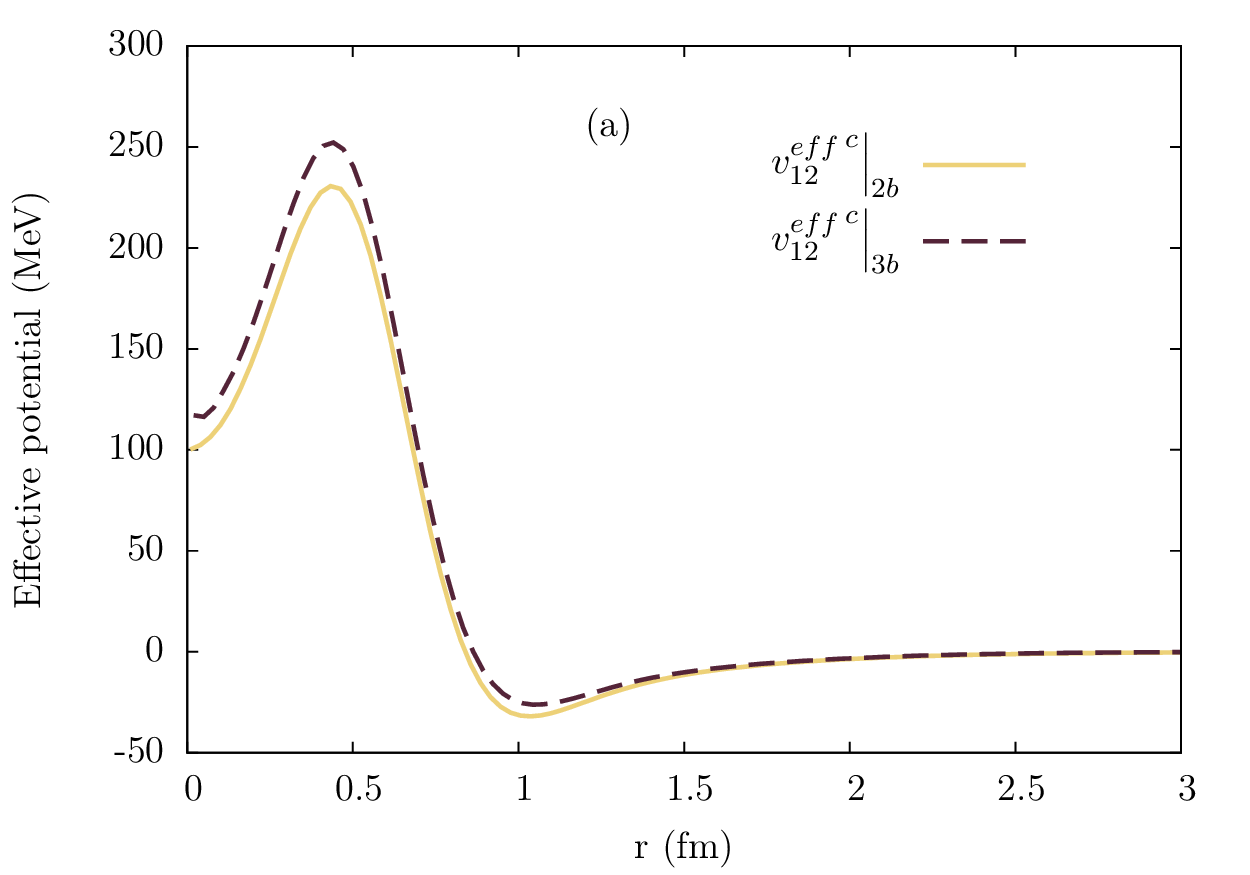}
\vspace{0.1cm}
\\
\includegraphics[width=8.0cm]{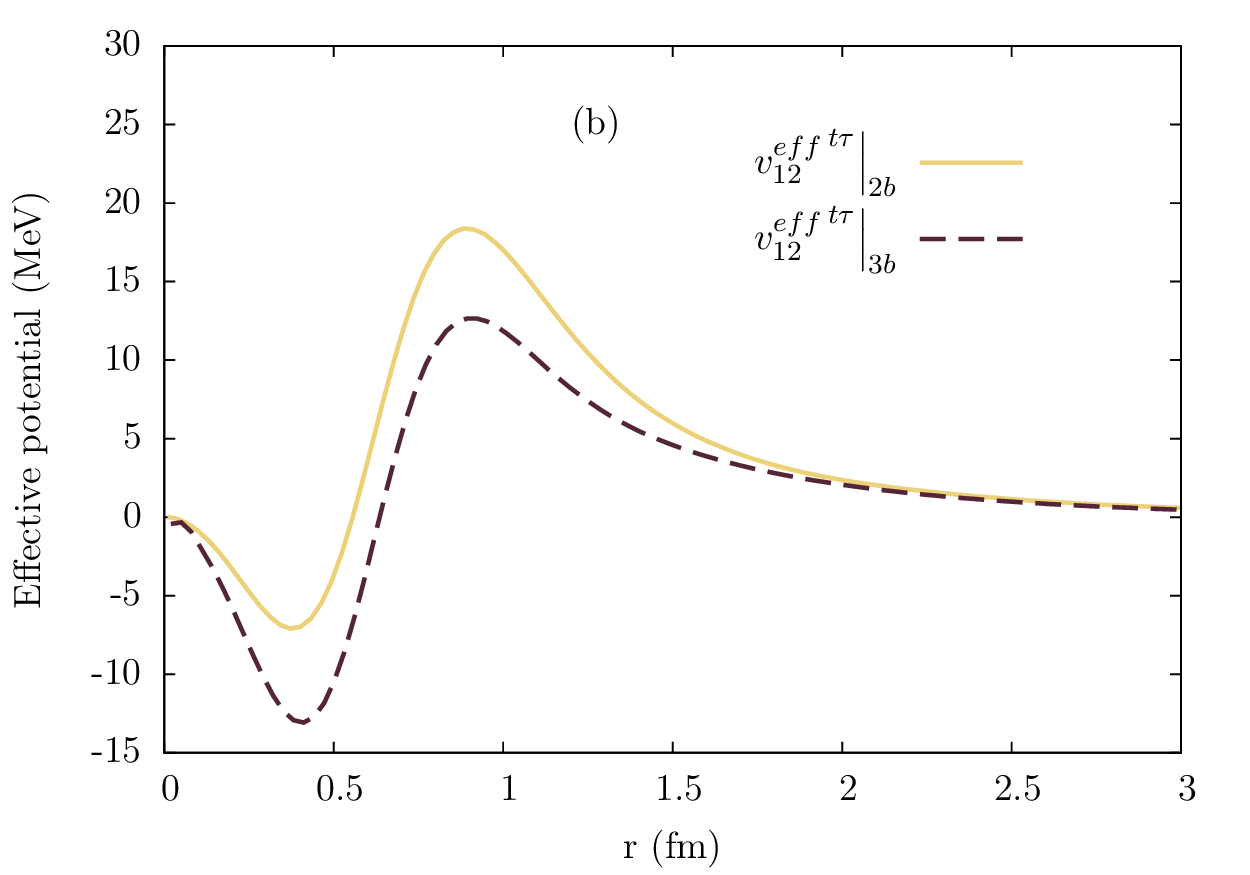}
\vspace{0.1cm}
\caption{Central (a) and the $t\tau$ (b) channels of the effective potentials at two-body and three-body cluster level calculated for SNM at $\rho=0.16\,\text{fm}^{-3}$. \label{fig:eff_pot_comp}}
\end{center}
\end{figure}

\begin{figure}[!ht]
\begin{center}
\includegraphics[width=8.5cm]{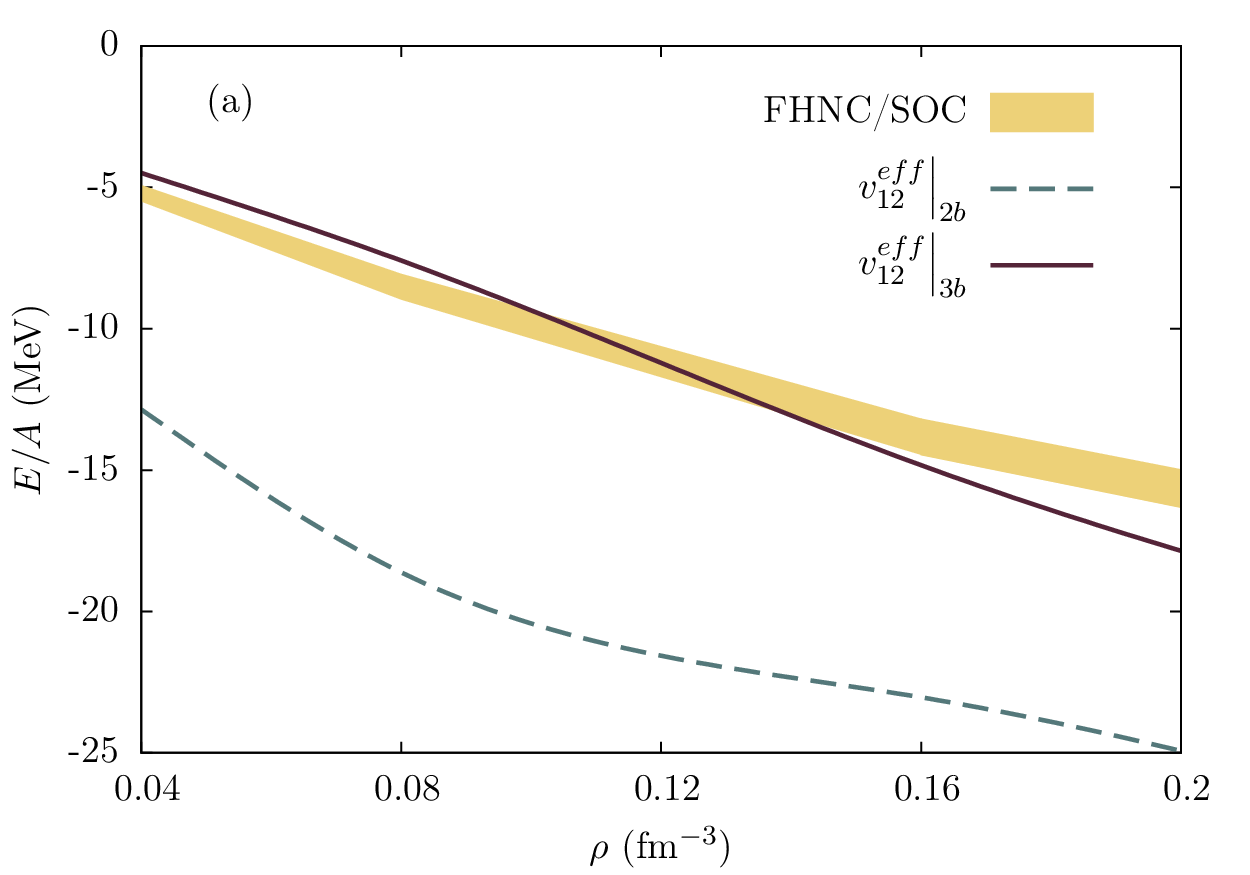}
\vspace{0.1cm}
\\
\includegraphics[width=8.5cm]{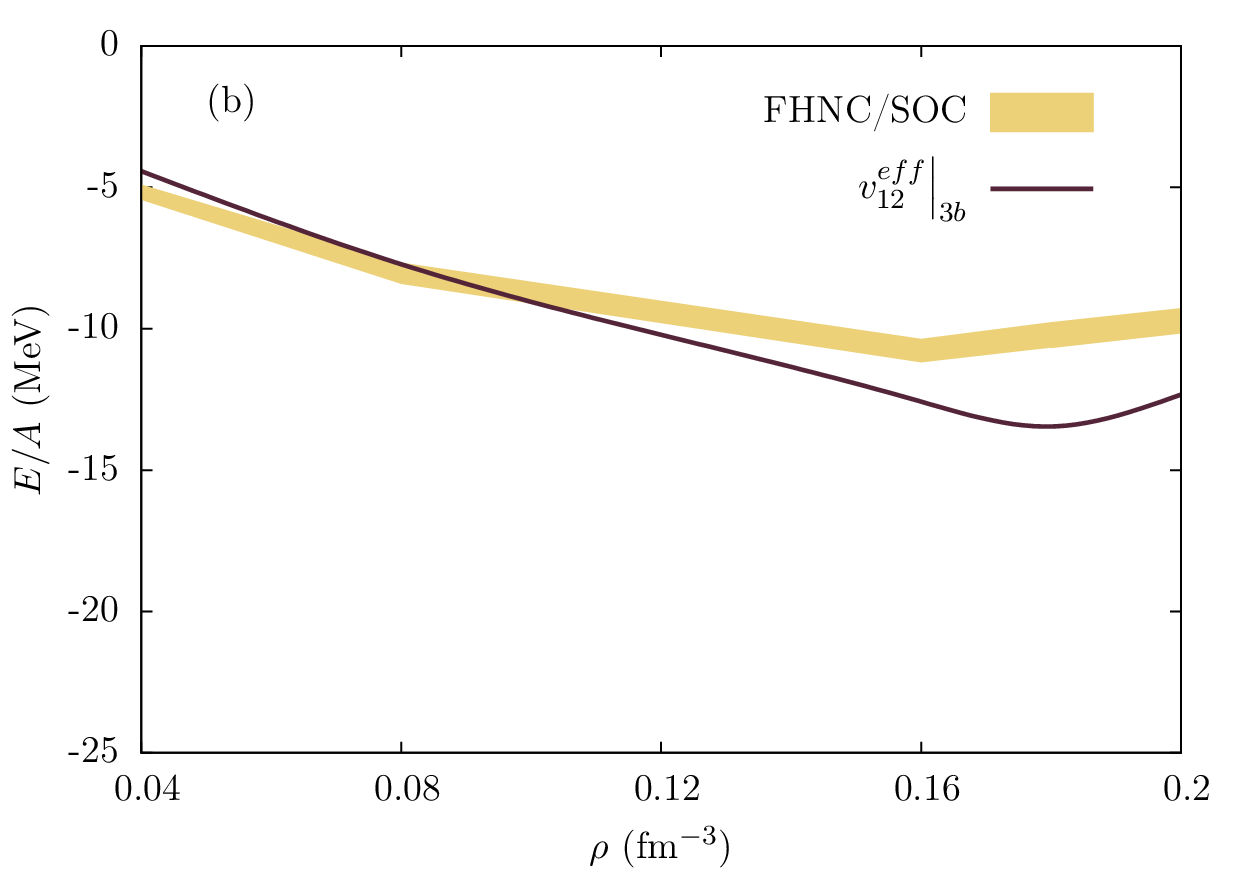}
\vspace{0.1cm}
\caption{EoS of SNM in the low density regime. In the panel (a) the bare hamiltonian only contains the Argonne $v_{6}^\prime$ potential, while in the panel (b) the UIX three-body interaction model is added to it. The dotted and the solid lines display the three-body and the two-body cluster effective potential results, respectively. The FHNC/SOC calculations are represented by the shaded region, accounting for the PB and JF kinetic energy difference. \label{fig:eos_eff}}
\end{center}
\end{figure}
Note that the above equation summarizes the terms corresponding to Eqs. (\ref{eq:v3b_dir}), (\ref{eq:v3b_p12}), (\ref{eq:v3b_p13}) and (\ref{eq:v3b_cir}). A comparison with Eq. (\ref{eq:eff_pot}) immediately leads to 
\begin{align}
\hat{v}_{12}^{eff}\Big|_{3b}=&\rho\int d\mathbf{r}_3 \text{CTr}_3 \Big[ \hat{X}(x_1,x_2;x_3)\Big(1-\hat{P}^{\sigma\tau}_{12}\ell^{2}_{12}\nonumber\\
&-2\hat{P}^{\sigma\tau}_{13}\ell^{2}_{13}+2\hat{P}^{\sigma\tau}_{12}\hat{P}^{\sigma\tau}_{13}\ell_{12}\ell_{13}\ell_{23}\Big)(1-\hat{P}_{12}^{\sigma\tau}\ell_{12}^2)^{-1}\Big]\, ,
\end{align}
where the subscript $3$ indicates that the spin-isospin normalized trace has to be performed over coordinates of particle $3$ only. A similar argument holds for the three-body cluster term of the kinetic energy and the three-body potential contributions to the effective potential. 

In Fig. \ref{fig:eff_pot_comp} the central and $t\tau$ components of the effective potentials at two-body and three-body cluster level are compared. The starting bare NN interaction is the Argonne $v_{6}^\prime$; for the three-body cluster results the UIX three-body potential has been included in the calculations. 

Starting from a bare hamiltonian only containing the Argonne $v_{6}^\prime$ NN potential, we have computed the EoS of SNM for the low-density regime using both the new three-body cluster effective interaction and the older one with only two-body cluster diagrams. The results have been compared with the corresponding FHNC/SOC calculations, displayed as a shaded region Fig. \ref{fig:eos_eff} to account for the PB and JF kinetic energy difference. The curve corresponding to $v_{12}^{eff}\big{|}_{3b}$ is much closer to the FHNC/SOC results than the one obtained with the older $v_{12}^{eff}\big{|}_{3b}$. 

In the lower panel of Fig. \ref{fig:eos_eff}, the EoS of SNM are shown for an hamiltonian containing the Argonne $v_{6}^\prime$ NN potential along with the UIX three-body interaction model. Again the curve obtained from the three-body cluster effective potential is close to the full calculation and, it exhibits the saturation, which is a remarkable feature.

As a second step in the definition of the effective interaction, we have adjusted the variational parameters of the correlation functions in order for the effective hamiltonian exactly reproduces the energy per particle obtained with the full FHNC/SOC calculations at saturation density. In particular the ``healing distance'', of both the central, $d_c$, and the tensorial, $d_t$, correlations and the quenching parameters $\alpha_p$ \cite{lovato_11}, which can be given the interpretation of the low-energy parameters of the effective interaction, have been reduced.

\section{Effective weak operators}
\label{sec:ewo}
In the case of $1p-1h$ excitation, the effective weak operators  $\hat{O}_{\textbf{q}}^{eff}$, introduced in Ref. \cite{cowell_04}, are defined through the relation
\begin{align}
&\langle \Phi_{p_m; h_i} | \hat{O}_{\mathbf{q}}^{eff}| \Phi_0 \rangle \equiv \frac{\langle \Phi_{p_m; h_i} | \mathcal{F}^\dagger \hat{O}_{\mathbf{q}}  \mathcal{F} | \Phi_0\rangle}{\sqrt{\langle \Phi_0 | \mathcal{F}^\dagger \mathcal{F} | \Phi_0\rangle\langle \Phi_{p_m; h_i} | \mathcal{F}^\dagger \mathcal{F} | \Phi_f\rangle}}
\label{eq:cbf_1p1heff}
\end{align}
 
For the sake of giving a unified description of matrix elements associated with both the Fermi and Gamow-Teller transitions, it is convenient to distinguish the common radial part from the specific zero momentum form factors and spin-isospin operators, defining
\begin{equation}
\hat{O}_{\mathbf{q}}(1)\equiv e^{i\mathbf{q}\cdot\mathbf{r}_1} \hat{O}_{\sigma\tau}(1)\ .
\end{equation}
From Eq. (\ref{eq:f_gt_def}) it follows that $\hat{O}_{\sigma\tau}^{F}(1)=g_V \tau^{+}_1$ and $\hat{O}_{\sigma\tau}^{GT}(1)=g_A\vec{\sigma}_1 \tau^{+}_1$. 
 
As for the calculation of the hamiltonian expectation value, a cluster expansion of the weak operator correlated matrix element can be performed \cite{cowell_03}. The smallness parameters in this case are $f^{c}_{ij}-1$ and $f^{p}_{ij}$. Following Ref.\cite{fantoni_87}, we denote by $c_i$ the quantum numbers of occupied states in both $\Phi_0$ and $\Phi_{p_m; h_i}$.  It can be shown \cite {fantoni_87,lovato_12c} that the CBF matrix element of the effective weak transition operator takes the form 
\begin{align}
&\langle \Phi_{p_m; h_i} | \hat{O}_{\mathbf{q}}^{eff}| \Phi_0 \rangle=\frac{1+\sum_C (q;p_m,h_i)}{\sqrt{1+\sum_C (h_i)}\sqrt{1+\sum_C (p_m)}}\, .
\label{eq:mat_el_sc}
\end{align}
The term $\sum_C (h_i)$ contains all the connected diagrams with one $h_i$ vertex or one $h_i$-exchange line. An analogous definition applies for $\sum_C (h_i)$. On the other hand, $\sum_C (q;p_m,h_i)$ amounts of connected diagrams having one single $p_mh_i$-exchange line or one single $p_mh_i$ vertex. Moreover, the weak operator $\hat{O}_{\mathbf{q}}(1)$, which carries a momentum $\mathbf{q}$ and a spin-isospin operator, is attached to the point $1$ of the diagrams belonging to $\sum_C (q;p_m,h_i)$. Note that only $c_i$ states are present in both the bare vertex and the bare exchange lines. In other words, unlike in the cluster expansion for the energy per particle, the hole state $h_i$ is lacking.

The two-body cluster diagrams coming from both the numerator and denominator of Eq. (\ref{eq:mat_el_sc}) have been computed by the authors of Ref. \cite{cowell_03}, while in Ref. \cite{benhar_09,farina_09} a different truncation scheme has been adopted and the numerator has been approximated to the unity. 

In this work, in addition to the numerator and denominator two-body cluster diagrams, for the sake of consistency with the effective interaction, we have calculated the three-body cluster diagrams at first order in $\hat{f}-1$. They are associated with the numerator of Eq. (\ref{eq:mat_el_sc}), as they originate from
\begin{equation}
\{\hat{f}_{23}-1, \hat{O}_{\mathbf{q}}(1)\}\,=2 \hat{O}_{\mathbf{q}}(1)(\hat{f}_{23}-1)\,.
\end{equation}

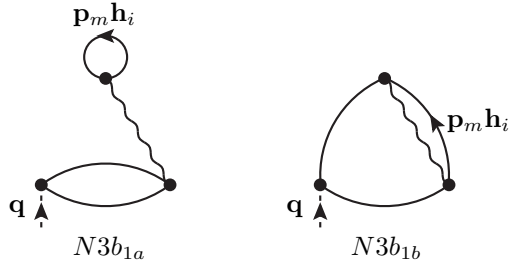
\begin{figure}[!t]
\begin{center}
  \begin{picture}(300,100) (0,-30)
    \SetWidth{1.0}
    \SetColor{Black}
    \SetScale{0.8}	
    \unitlength=0.8 pt
	\Photon(90,50)(120,0){1}{5}
	\Vertex(60,0){3}
       	\Vertex(90,50){3}
        \Vertex(120,0){3}
        	\Arc(90,-40)(50,50,130)
	\Arc(90,40)(50,230,310)
	\ArrowArc(90,60)(10,-90,270)
	\DashArrowLine(60,-20)(60,-5){3}
	\Text(45,-15)[lb]{$\mathbf{q}$}
	\Text(75,75)[lb]{$\mathbf{p}_m\mathbf{h}_i$}
	\Text(75,-35)[lb]{$N3b_{1a}$}
	\Photon(220,50)(250,0){1}{5}
	\ArrowArc(200,4.43)(50,-5,65)
	\Arc(240,4.43)(50,115,185)
	\Arc(220,40)(50,230,310)
	\Vertex(190,0){3}
       	\Vertex(220,50){3}
        \Vertex(250,0){3}
        \DashArrowLine(190,-20)(190,-5){3}
	\Text(175,-15)[lb]{$\mathbf{q}$}
	\Text(250,25)[lb]{$\mathbf{p}_m\mathbf{h}_i$}
	\Text(205,-35)[lb]{$N3b_{1b}$}
	
\end{picture}
\caption{Non vanishing three body diagrams at first order in $\hat{f}-1$ emerging from the numerator of Eq. (\ref{eq:mat_el_sc}).\label{fig:num_3b_g_1}}
\end{center}
\end{figure}

\begin{figure}[!h]
\begin{center}
  \begin{picture}(100,100) (-20,-40)
    \SetWidth{1.0}
    \SetColor{Black}
    \SetScale{0.8}	
    \unitlength=0.8 pt
	\Photon(0,35)(65,35){1}{5}
	\DashArrowLine(0,15)(0,30){3}
	\Text(-15,15)[lb]{$\mathbf{q}$}
	\Vertex(0,35){3}
	\Vertex(65,35){3}
	\ArrowArc(65,45)(10,270,-90)	
	\Text(50,60)[lb]{$\mathbf{p}_m\mathbf{h}_i$}
	\Text(25,-5)[lb]{$N2b_{1a}$}
  \end{picture}
\end{center}    
\vspace{-1.5cm}
\caption{Two-body diagram of the first order term in $\hat{f}-1$ coming from the numerator of Eq. (\ref{eq:mat_el_sc}).}
\label{fig:2b_num}
\end{figure}
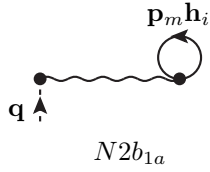

The only non vanishing three-body cluster diagrams are depicted in Fig. \ref{fig:num_3b_g_1}. In the thermodynamic limit, diagram $N3b_{1a}$ reads
\begin{align}
N3b_{1a}=&-\frac{2\rho^2}{\nu^2}\delta_{\mathbf{q},\mathbf{p}_m-\mathbf{h}_i}\int d\mathbf{r}_{12}e^{i\mathbf{q}\cdot\mathbf{r}_{12}}\ell^2_{12} \int d\mathbf{r}_{23}e^{i\mathbf{q}\cdot\mathbf{r}_{23}}\nonumber\\
&\times\sum_{\alpha_i}\langle \alpha_1\alpha_2 \alpha_{p_m}|\hat{O}_{\sigma\tau}(1) (\hat{f}_{23}-1)\hat{P}^{\sigma\tau}_{12}|\alpha_1\alpha_2 \alpha_{h_i}\rangle\, ,
\end{align}
where $|\alpha_i\rangle$ denotes the spin-isospin state of particle $i$. The discretized momentum conservation is expressed by the Kronecker delta function. By computing the spin-isospin matrix elements, whose values are reported in Appendix E of Ref. \cite{lovato_12c}, one can show that in the limit of zero momentum transfer, $\mathbf{q}\to 0$, $N3b_{1a}$ cancels the contribution of diagram $N2b_{1a}$, represented in Fig. \ref{fig:2b_num} (denoted as F1d j and GT1d j in Ref. \cite{cowell_03}). This is an indication that three-body diagrams need to be taken into account and they play a relevant role in the sum rules of the weak response, which will be estimated at a later stage.

It can be readily shown that the analytic expression of diagram $N3b_{1b}$ is given by 
\begin{align}
N3b_{1b}=&\frac{2\rho^2}{\nu^2}\delta_{\mathbf{q},\mathbf{p}_m-\mathbf{h}_i}\int d\mathbf{r}_{12}d\mathbf{r}_{23}e^{i \mathbf{p}_m\cdot\mathbf{r}_{13}}e^{-i \mathbf{h}_i\cdot\mathbf{r}_{12}}\ell_{12}\ell_{13}\nonumber\\
&\times\sum_{\alpha_i}\langle \alpha_1\alpha_2 \alpha_{p_m}|\hat{O}_{\sigma\tau}(1) (\hat{f}_{23}-1)\hat{P}^{\sigma\tau}_{12}\hat{P}^{\sigma\tau}_{13}|\alpha_1\alpha_2 \alpha_{h_i}\rangle\,.
\end{align}
The evaluation of the spin-isospin matrix elements can again be found in Appendix E of Ref. \cite{lovato_12c}.

\section{Response functions in different approximations}
\label{sec:resp_appr}
\subsection{Correlated Fermi gas (CFG) and correlated Hartree-Fock (CHF)} 
In both the correlated Fermi gas (CFG) and correlated Hartree-Fock (CHF) approximations, the weak response of cold SNM, defined in Eq. (\ref{eq:resp_def}), is given by \cite{cowell_04}
\begin{align}
S_{FG}(\mathbf{q},\omega)=&\frac{1}{A}\sum_{p_mh_i} |\langle \Phi_{p_m;h_i} | \hat{O}_{\mathbf{q}}^{eff}| \Phi_0 \rangle|^2 \delta(\omega+\epsilon_{p_m}-\epsilon_{h_i})\, .
\label{eq:cfg}
\end{align}
Within the CFG approximation, the single particle energies are those of the non-interacting hamiltonian
\begin{equation}
\epsilon_{n_i}=\frac{\mathbf{k}_{i}^2}{2m}\, .
\label{eq:spe_fg}
\end{equation}

The single particle approximation is retained in the CHF; however the potential enter the calculation of the single particle energies. It has long known \cite{jastrow_55} that the Hartree-Fock approximation is not suitable for nuclear potentials having a repulsive core, like the Argonne models, because it does not encompass the correlations between nucleons. We use 
instead the effective potential described in Section \ref{sec:ei}, which is appropriate for mean field calculations. The single particle energy is then given by
\begin{align}
\epsilon_{n_i}=&\frac{\mathbf{k}_{i}^2}{2m}+\sum_{n_j=1}^A \int dx_{j}\phi_{n_i}^*(x_i)\phi_{n_j}^*(x_j)v_{ij}^{eff}\mathcal{A}[\phi_{n_i}(x_i)\phi_{n_j}(x_j)]\, ,
\end{align}
where $\mathcal{A}$ is the antisymmetrizing operator.
The single particle wave functions are plane waves
\begin{equation}
\phi_{n_i}(x_i)=\frac{e^{i\mathbf{k}_i \cdot \mathbf{r}_i}}{\sqrt{V}}\eta_{\alpha_i}\, ,
\label{eq:nm_wf}
\end{equation}
where $\eta_{\alpha_i}\equiv\chi_{\sigma_i}\chi_{\tau_i}$ represents the product of Pauli spinors describing the spin and the isospin of particle $i$ and $V$ is the normalization volume. 

While in CFG calculations the correlations enter only through the effective weak operators, within the CHF approximations they sizably determine the effective interaction. 

In the case of SNM ($\nu=4$) for potentials of the form of Argonne $v_{18}$, carrying out the summation over the occupied states with $|\mathbf{k}_j|\leq k_F$ yields
\begin{align}
\epsilon_{n_i}=&\frac{\mathbf{k}_{i}^2}{2m}+\rho \int d\mathbf{r}_{ij}\Big[v^{c}_{ij}-\frac{1}{4}\ell(k_Fr_{ij})e^{-i\mathbf{k}_i\cdot \mathbf{r}_{ij}}(v^{c}_{ij}+3v^{\tau}_{ij}+3v^{\sigma}_{ij}+9v^{\sigma\tau}_{ij})\Big]\,.
\label{eq:spe_nm2}
\end{align}
For the sake of simplicity, in the latter equation the superscript ``${eff}$'' has been omitted.

\subsection{Correlated Tamm-Dancoff (CTD)}
Since the correlated $1p-1h$ states are not eigenstates of the full nuclear hamiltonian, transitions between them are in principle allowed. They can be accounted for within the Tamm-Dancoff approximation, that amounts to expanding the final state of Eq. (\ref{eq:resp_def}) in series of $1p-1h$ excitations.

Because the hamiltonian is translationally invariant, the total momentum $\mathbf{q}$ of the state is conserved, and the momenta of the particle, $\mathbf{p}_m$, and the hole, $\mathbf{h}_i$, satisfy the relation $\mathbf{p}_m-\mathbf{h}_i=\mathbf{q}$.

The nuclear hamiltonian commutes with the total isospin, $T$  with the total isospin projection along the $z-$axis, $T_z$, and with the total spin, $S$. However, because of the tensor term of the potential, the hamiltonian does not commute with $S_z$, the total spin projection along the $z-$ axis. 

The combinations of particle hole pairs that are eigenstates of $S$ and $S_z$, that define the particle-hole Clebsch-Gordan coefficients, are shown in Table \ref{tab:s_ph}. The differences between the total spin states of the particle particle pairs, also given in Table \ref{tab:s_ph}, are due to the phase factor appearing in the canonical transformations to particles and holes \cite{fetter_03}. The treatment of the total isospin can be done in complete analogy, replacing the up and the down single particle spin states with the proton and the neutron isospin states, respectively. 
\begin{table}[h!]
\begin{center}
\renewcommand*\arraystretch{1.5}
\caption{Spin configurations for a particle particle pair
and a particle hole pair for spin-1/2 particles. \label{tab:s_ph}}
\vspace{0.3cm}
\begin{tabular}{l c c }
\hline
\hline
Total spin state &    particle particle      &  particle hole \\
\hline
$S=1\,,S_z=1$  & $\uparrow\uparrow$ & $-\uparrow\downarrow$ \\
$S=1\,,S_z=0$  & $\frac{1}{\sqrt{2}}(\uparrow\downarrow+\downarrow\uparrow)$ & $\frac{1}{\sqrt{2}}(\uparrow\uparrow-\downarrow\downarrow)$  \\
$S=1\,,S_z=-1$ & $\downarrow\downarrow$ & $\downarrow\uparrow$ \\
$S=0\,,S_z=0$  & $\frac{1}{\sqrt{2}}(\uparrow\downarrow-\downarrow\uparrow)$ & $\frac{1}{\sqrt{2}}(\uparrow\uparrow+\downarrow\downarrow)$  \\

\hline
\hline
\end{tabular}
\vspace{0.1cm}
\end{center}
\end{table}    

It is possible to reduce the computational effort necessary for solving the Tamm-Dancoff equations, by starting with combinations of $|\Phi_{p_m;h_i}\rangle$ with definite $T$, $T_z$, $S$ and $S_z$
\begin{equation}
|\Phi_n\rangle_{TT_zS}^{TDA}=\sum_{\mathbf{p}_m\mathbf{h}_i S_z}C^{n\,TT_zSS_z}_{\mathbf{p}_m\mathbf{h}_i} |\Phi_{\mathbf{p}_m;\mathbf{h}_i}\rangle_{TT_zSS_z}\, ,
\end{equation}
A further simplification arises by noting that the final states of both the Fermi and the Gamow-Teller transitions are characterized by having $T=1$ and $T_z=1$. To simplify the notation, the  isospin indexes may then be omitted
\begin{equation}
|\Phi_n\rangle_{S}^{TDA}=\sum_{\mathbf{p}_m\mathbf{h}_i S_z}C^{n\,SS_z}_{\mathbf{p}_m\mathbf{h}_i} |\Phi_{\mathbf{p}_m;\mathbf{h}_i}\rangle_{SS_z}\, ,
\label{eq:TDA_exp_new}
\end{equation}
and it is understood that $T=1$ and $T_z=1$.

The eigenvalue equation for the effective hamiltonian defines the excitation energy $\omega_{n}^S$
\begin{align}
\hat{H}^{eff}|\Phi_n\rangle^{TDA}_S=&\Big(\sum_i-\frac{\nabla^{2}_i}{2m}+\sum_{i<j}\hat{v}^{eff}_{ij}\Big)|\Phi_n\rangle^{TDA}_S=(E_0+\omega_{n}^S)|\Phi_n\rangle^{TDA}_S\, .
\label{eq:eigen_TDA}
\end{align}
Multiplying from the left the previous equation by $\:_{SS_z}\langle\Phi_{\mathbf{p}_n;\mathbf{h}_j}|$ and using the orthonormality of the $1p-1h$ states yields
\begin{align}
&\sum_{\mathbf{p}_m\mathbf{h}_i S'_z}\:_{SS_z}\langle\Phi_{\mathbf{p}_n;\mathbf{h}_j}|\hat{H}^{eff}|\Phi_{\mathbf{p}_m;\mathbf{h}_i}\rangle_{SS'_z}\,C^{n\,SS'_z}_{\mathbf{p}_m\mathbf{h}_i} =C^{n\,SS_z}_{\mathbf{p}_n\mathbf{h}_j}(E_0+\omega_{n}^{S})\, .
\end{align}
Thus, finding the coefficient $C^{n\,SS_z}_{\mathbf{p}_m\mathbf{h}_i} $ amounts in diagonalizing the block diagonal hamiltonian for the two subsets of the $1p-1h$ basis having $T=1$, $T_z=1$ corresponding to $S=0$ and to $S=1$ . This is a much less expensive computational task than diagonalizing the hamiltonian in the full $1p-1h$ basis. As a consequence, this approach allows for considering a larger number of momentum states.

It has been shown \cite{lovato_12c} that, singling out particles $1$ and $2$ from both the ground state and the $1p-1h$ excited state, the matrix element of the hamiltonian reads
\begin{align}
\:_{SS_z}\langle\Phi_{\mathbf{p}_n;\mathbf{h}_j}|\hat{H}^{eff}|\Phi_{\mathbf{p}_m;\mathbf{h}_i}\rangle_{SS'_z}=&\Big[(E_0+\epsilon_{\mathbf{p}_m}-\epsilon_{\mathbf{h}_i})\delta_{\mathbf{p}_m\mathbf{p}_n}\delta_{\mathbf{h}_i\mathbf{h}_j}\delta_{S_zS'_z}\nonumber\\
&+\langle \mathbf{p}_n\,\mathbf{h}_i|\hat{v}^{eff}_{12}|\mathbf{h}_j\,\mathbf{p}_m\rangle_{SS'_zS_z}\Big]\label{eq:eigen_TDA5}
\end{align}

As far as the notation is concerned, with $\langle p_n\,h_j|O_{12}|h_i\,p_m\rangle$ we denote the two-body matrix element of the operator $\hat{O}_{12}$ 
\begin{align}
&\langle p_n\,h_i|\hat{O}_{12}|h_j\,p_m\rangle\equiv\int dx_{1,2} \phi_{p_n}^*(x_1)\phi_{h_i}^*(x_2)\hat{O}_{12}\mathcal{A}[ \phi_{h_j}(x_1)\phi_{p_m}(x_2)]\, .
\end{align}

In the two-body matrix element of the effective potential $\langle \mathbf{p}_n\,\mathbf{h}_i|\hat{v}^{eff}_{12}|\mathbf{h}_j\,\mathbf{p}_m\rangle_{SS'_zS_z}$ the spin and isospin projections along the $z$ axis of the particle hole pairs $p_m-h_i$ and $p_n-h_j$ are combined as in Table \ref{tab:s_ph} to have definite $S$,  and total spin projections along $z$ equal to $S'_z$ and $S_z$, respectively. We 
recall that the total isospin and its $z$-projection are $T=1$ and $T_z=1$.

The direct and exchange terms for the case $S=0$, relevant to the Fermi transition, for SNM read
\begin{align}
&\langle \mathbf{p}_n\,\mathbf{h}_i|\hat{v}^{eff}_{12}|\mathbf{h}_j\,\mathbf{p}_m\rangle_{000}^{d}= \frac{4}{V}\int d\mathbf{r}_{12}e^{-i\mathbf{q}\cdot\mathbf{r}_{12}}\,v_{12}^\tau 
\nonumber\\
&\langle \mathbf{p}_n\,\mathbf{h}_i|v_{12}|\mathbf{h}_j\,\mathbf{p}_m\rangle_{000}^{e}=\frac{1}{V}\int d\mathbf{r}_{12}e^{i\mathbf{k}_{ij}\cdot\mathbf{r}_{12}}(v^c_{12} - v^\tau_{12} + 3 v^\sigma_{12} - 3v^{\sigma\tau}_{12})\, ,
\end{align}
where $\mathbf{k}_{ij}\equiv\mathbf{h}_i-\mathbf{h}_j$. Again, for the sake of simplicity, the superscript ``${eff}$'' has been omitted where the channels of the effective potential are specified.

For the Gamow Teller transition, the final state has $S=1$; hence it is necessary to compute the nine matrix elements $\langle \mathbf{p}_n\,\mathbf{h}_i|\hat{v}^{eff}_{12}|\mathbf{h}_j\,\mathbf{p}_m\rangle_{SS'_zS_z}$ corresponding to $S_z,S'_z=-1\,,0\,,1$

\begin{align}
\langle \mathbf{p}_n\,\mathbf{h}_i|\hat{v}^{eff}_{12}|\mathbf{h}_j\,\mathbf{p}_m\rangle_{111}^{d}=&\frac{2}{V}\int d\mathbf{r}_{12}e^{-i\mathbf{q}\cdot\mathbf{r}_{12}}\Big[
2v^{\sigma\tau}_{12}+v^{t\tau}_{12}\Big(1-\frac{3z_{12}^2}{r_{12}^2}\Big)\Big]\nonumber\\
\langle \mathbf{p}_n\,\mathbf{h}_i|\hat{v}^{eff}_{12}|\mathbf{h}_j\,\mathbf{p}_m\rangle_{111}^{e}=&\frac{1}{V}\int d\mathbf{r}_{12}e^{i\mathbf{k}_{ij}\cdot\mathbf{r}_{12}}\Big[v^{c}_{12} - v^{\tau}_{12} - v^{\sigma}_{12} + v^{\sigma\tau}_{12}\nonumber\\
&+(v^{t}_{12}-v^{t\tau}_{12})\Big(1-3\frac{z_{12}^2}{r_{12}^2}\Big)\Big]\nonumber\\
\langle \mathbf{p}_n\,\mathbf{h}_i|\hat{v}^{eff}_{12}|\mathbf{h}_j\,\mathbf{p}_m\rangle_{101}^{d}=&-\frac{6\sqrt{2}}{V}\int d\mathbf{r}_{12}e^{-i\mathbf{q}\cdot\mathbf{r}_{12}}\:v^{t\tau}_{12}
\frac{(x_{12}-iy_{12})z_{12}}{r_{12}^2}\nonumber\\
\langle \mathbf{p}_n\,\mathbf{h}_i|\hat{v}^{eff}_{12}|\mathbf{h}_j\,\mathbf{p}_m\rangle_{101}^{e}=&-\frac{3\sqrt{2}}{V}\int d\mathbf{r}_{12}e^{i\mathbf{k}_{ij}\cdot\mathbf{r}_{12}}\:(v^{t}_{12}-v^{t\tau}_{12})
\frac{(x_{12}-iy_{12})z_{12}}{r_{12}^2}\nonumber\\
\langle \mathbf{p}_n\,\mathbf{h}_i|\hat{v}^{eff}_{12}|\mathbf{h}_j\,\mathbf{p}_m\rangle_{1-11}^{d}=&-\frac{6}{V}\int d\mathbf{r}_{12}e^{-i\mathbf{q}\cdot\mathbf{r}_{12}}\:v^{t\tau}_{12}\frac{(x_{12}-iy_{12})^2}{r_{12}^2}\nonumber\\
\langle \mathbf{p}_n\,\mathbf{h}_i|\hat{v}^{eff}_{12}|\mathbf{h}_j\,\mathbf{p}_m\rangle_{1-11}^{e}
=&-\frac{3}{V}\int d\mathbf{r}_{12}e^{i\mathbf{k}_{ij}\cdot\mathbf{r}_{12}}\:(v^{t}_{12}-v^{t\tau}_{12})
\frac{(x_{12}-iy_{12})^2}{r_{12}^2}\nonumber\\
\langle \mathbf{p}_n\,\mathbf{h}_i|\hat{v}^{eff}_{12}|\mathbf{h}_j\,\mathbf{p}_m\rangle_{100}^{d}
=&\frac{4}{V}\int d\mathbf{r}_{12}e^{-i\mathbf{q}\cdot\mathbf{r}_{12}}\:\Big[v^{\sigma\tau}_{12}-v^{t\tau}_{12}\Big(
1-3\frac{z_{12}^2}{r_{12}^2}\Big)\Big]\nonumber\\
\langle \mathbf{p}_n\,\mathbf{h}_i|\hat{v}^{eff}_{12}|\mathbf{h}_j\,\mathbf{p}_m\rangle_{100}^{e}=&\frac{1}{V}\int d\mathbf{r}_{12}e^{i\mathbf{k}_{ij}\cdot\mathbf{r}_{12}}\:\Big[v^{c}_{12}-v^{\tau}_{12}-v^{\sigma}_{12}+v^{\sigma\tau}_{12}\nonumber\\
&-2(v^{t}_{12}-v^{t\tau}_{12})\Big(1-3\frac{z_{12}^2}{r_{12}^2}\Big)\Big]\,.
\end{align}

Replacing the final state of Eq. (\ref{eq:TDA_exp_new}) in the definition of the response, Eq. (\ref{eq:resp_def}) yields
\begin{align}
S(\mathbf{q},\omega)=&\frac{1}{A}\sum_n\Big{|}\sum_{\mathbf{p}_m\mathbf{h}_i S_z}C^{n\,SS_z}_{\mathbf{p}_m\mathbf{h}_i}\,_{SS_z}\langle \Psi_{\mathbf{p}_m;\mathbf{h}_i}| \hat{O}_\mathbf{q} |\Psi_0\rangle\Big{|}^2\delta(\omega-\omega_{n}^S)\, .
\end{align}

\section{Numerical calculation of the response}
\label{sec:num_calc}
We model the infinite system using a cubic box of side $L=V^{1/3}$ with periodic boundary conditions. Hence, the single particle wave functions are the plane waves of Eq. (\ref{eq:nm_wf}) with the discrete momenta
\begin{equation}
\mathbf{k}=\frac{2\pi}{L}(n_{k_x},n_{k_y},n_{k_z})  \, .\qquad n_{k_i}=0,\pm 1,\pm 2,\dots  \, .
\label{eq:discrete_mom}
\end{equation}
For zero temperature SNM, all single-particle states with $|\mathbf{k}|\leq k_F$ are occupied in the ground state. The momenta of the $ 1p-1h$ excitations are such that $|\mathbf{h}_j|\leq k_F$ and $|\mathbf{p}_m=\mathbf{h}_i+\mathbf{q}|> k_F$. For the hole and particle momentum to be on the lattice of allowed momentum states in the box, the momentum transfer must be such that 
\begin{equation}
\mathbf{q}=\frac{2\pi}{L}(n_{q_x},n_{q_y},n_{q_z})  \, .\qquad n_{q_i}=0,\pm 1,\pm 2,\dots  \, .
\end{equation}

Given the magnitude and the direction of the momentum transfer, the side of the box can be determined by inverting the latter equation
\begin{equation}
L=\frac{2\pi}{|\mathbf{q}|}\sqrt{n_{q_x}^2+n_{q_y}^2+n_{q_z}^2}\,.
\end{equation}

The size of the basis, that can be increased by increasing $\sqrt{n_{q_x}^2+n_{q_y}^2+n_{q_z}^2}$, has been determined requiring that the response of a system of noninteracting nucleons computed on the lattice agreed with the analytical result of the FG model \cite{cowell_04,benhar_09}. The FG response is obtained replacing the effective operator with the bare operator in Eq. (\ref{eq:cfg}) and using the single particle energy of Eq. (\ref{eq:spe_fg}). 

\begin{figure}[!h]
\begin{center}
\includegraphics[width=7.5cm,angle=0]{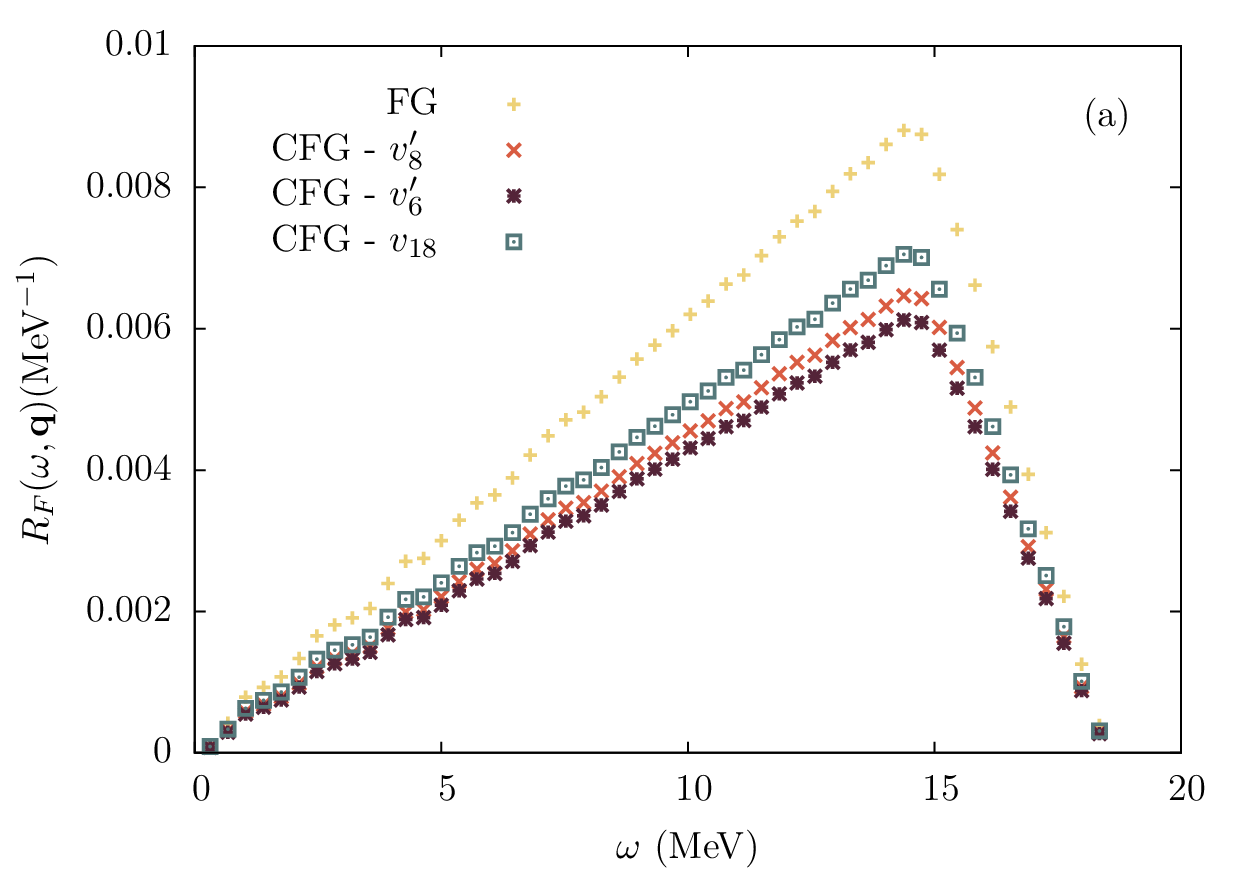}
\includegraphics[width=7.5cm,angle=0]{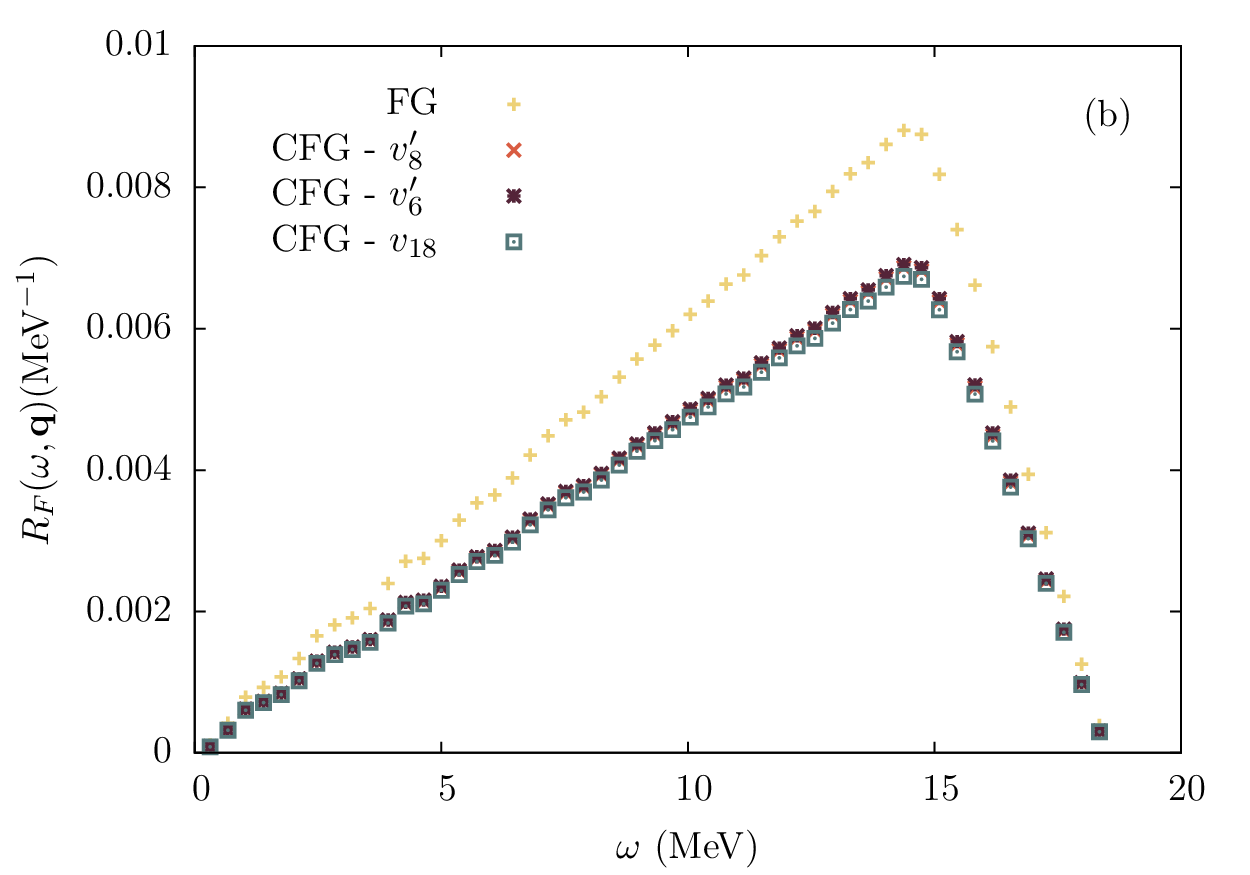}
\end{center}

\caption{Fermi response functions calculated at $q=0.3$ fm$^{-1}$. Panel (a) shows the response calculated at two-body cluster level for both $O_{\boldsymbol q}^{eff}$ and $v_{12}^{eff}$ for different choices of correlation functions. In panel (b), the three body cluster has been included. The plus marks show the response for a non interacting FG. \label{fig:cfg_chf_fermi}}
\end{figure}
The analytical calculations can be performed using a continuum of momentum states \cite{fetter_03}, while the numerical result consists of collection of discrete delta function peaked at the values of the single particle energies.  For a better representation of the results, as well as for fitting purposes, a gaussian representation of the energy conserving delta function has been adopted 
\begin{equation}
\delta(x-x_0)\to \frac{1}{\sigma\sqrt{\pi}}\exp\Big[-\Big(\frac{x-x_0}{\sigma}\Big)^2\Big]\, .
\end{equation}
For sufficiently small values of the gaussian width $\sigma$, the results become insensitive to it.
All the results that will be shown in this section refer to SNM at equilibrium density $\rho=0.16$ fm$^{-3}$.

\begin{figure}[!h]
\begin{center}
\includegraphics[width=7.5cm,angle=0]{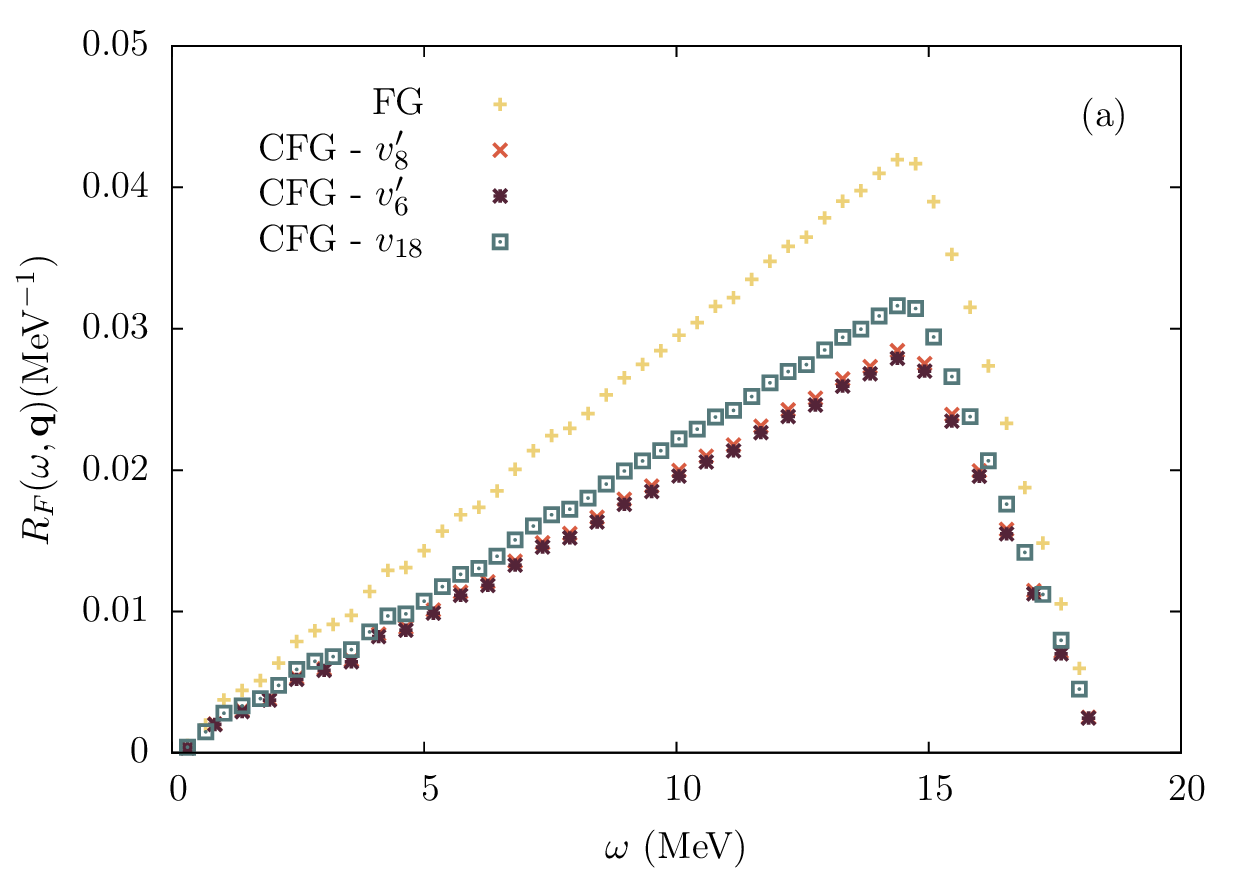}
\includegraphics[width=7.5cm,angle=0]{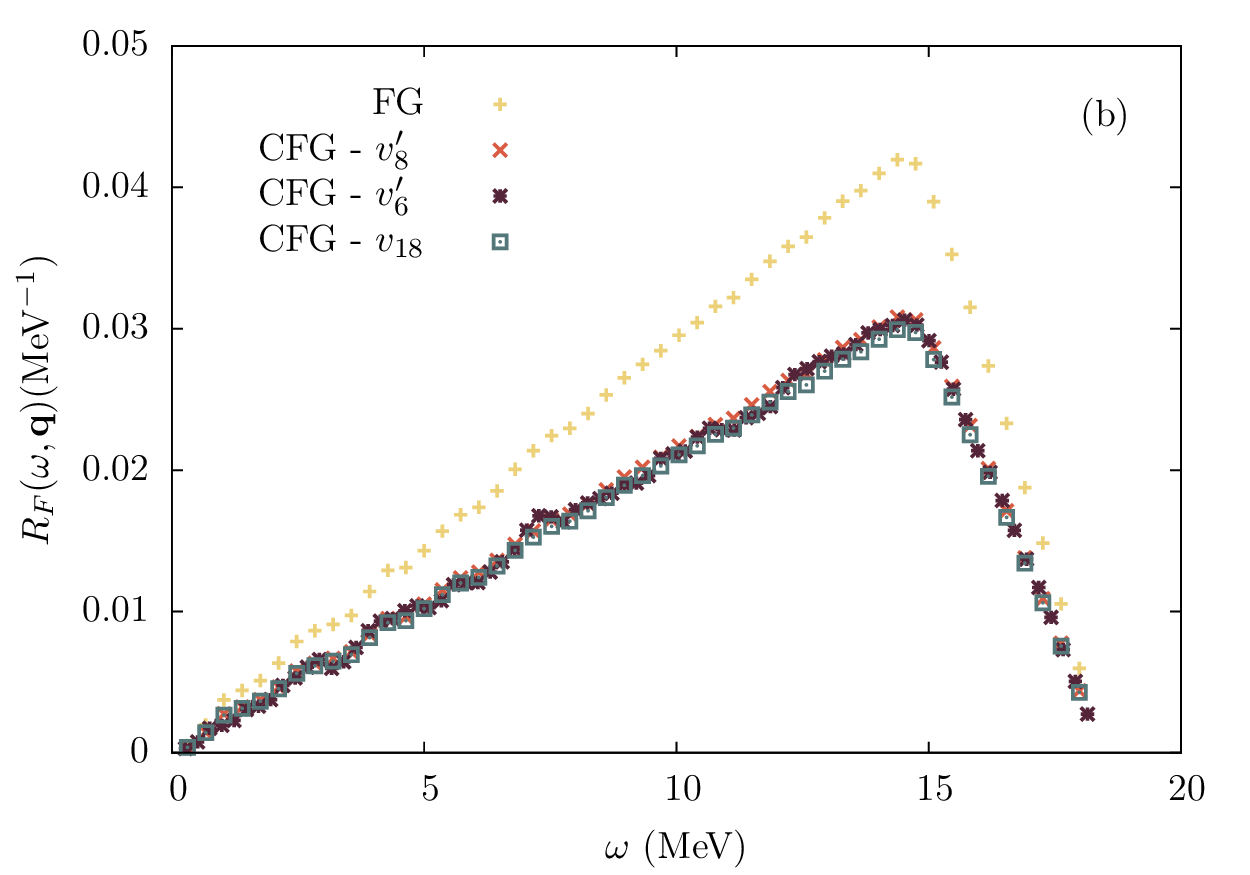}
\end{center}
\caption{Same as Fig. \ref{fig:cfg_chf_fermi} but for Gamow-Teller response functions. \label{fig:cfg_chf_gt}}
\end{figure}

\subsection{CFG}
FHNC/SOC  calculations and their associated minimization procedure provide a set of correlations function, corresponding to the minimum of the hamiltonian  expectation value. We have found the best correlation functions for the Argonne $v_{6}^\prime$, $v_{8}^\prime$ two-body potentials, and for comparison we have also considered the correlations of Ref. \cite{akmal_98} corresponding to Argonne $v_{18}$. With these correlations, the Fermi and Gamow-Teller response functions have been evaluated in CFG approximations with the sake of testing the contribution of the three-body cluster in the effective weak transition operators. The effective potential indeed does not enter the CFG calculations.

When only two-body cluster diagrams are considered, as in Refs. \cite{cowell_04,benhar_09}, the CFG response, suppressed by a $20-25\%$ with respect to the FG case, exhibits a sizable dependence on the choice of correlations, as shown in the panels (a) of Figs. \ref{fig:cfg_chf_fermi} and \ref{fig:cfg_chf_gt}. These figures refer to a transfer momentum 
$\boldsymbol q=|{\bf q}|(4\hat{x}+4\hat{y}+4\hat{z})/\sqrt{48}$ with $|\mathbf{q}|=0.3$ fm$^{-1}$. The folding gaussian function has a width of $0.25$ MeV. 

This unphysical effect is removed once the effective weak transition operator is computed at three-body cluster level. As a matter of fact, the CFG curves in the panels (b) of the aforementioned figures are very close, when not superimposed, to each other. Therefore our results appear to be more robust than those of Refs. \cite{cowell_04,benhar_09}, as the physical quantities should not  be sensitive to the details of the short range behavior of the correlation functions.

\subsection{CHF and CTD}


\begin{figure}[!!h]
\begin{center}
\includegraphics[width=7.9cm,angle=0]{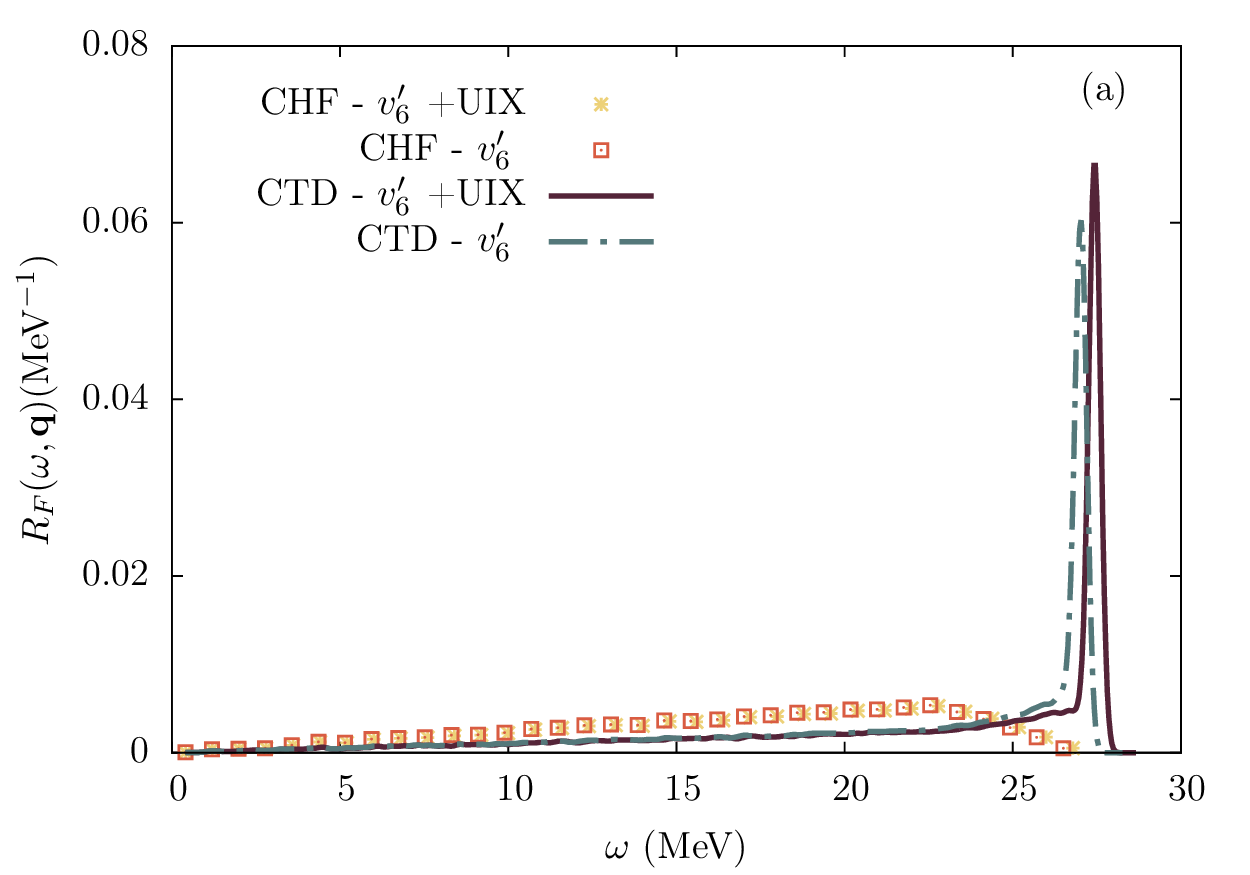}
\includegraphics[width=7.9cm,angle=0]{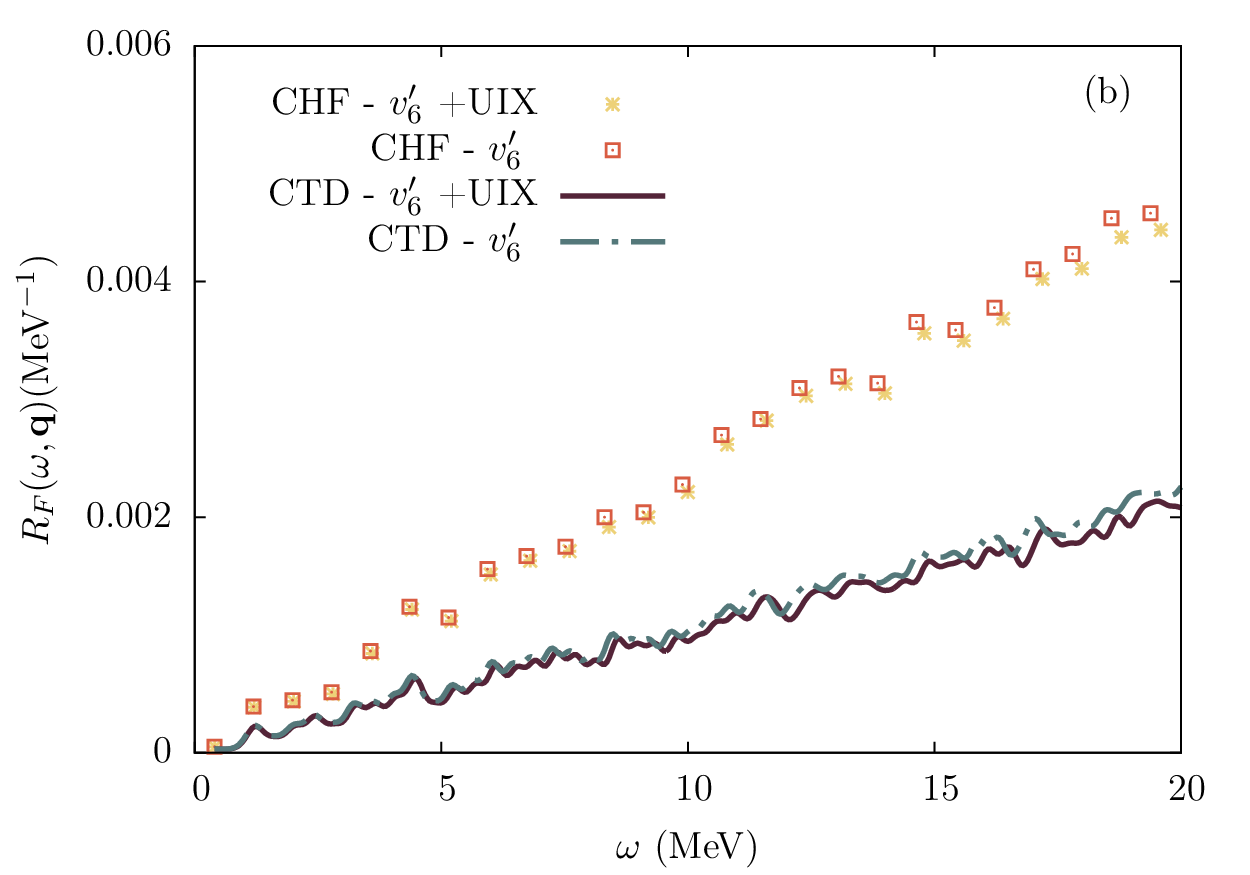}
\end{center}
\caption{Fermi response functions calculated at $q=0.3$ fm$^{-1}$ using $v_{6'}$ and $v_{6'}+UIX$ interaction models. Panel (a) shows the full response across all the values of $\omega$; panel (b) is a magnification of the small $\omega$ region. Responses are folded with a Gaussian of width 0.25 MeV.\label{fig:cfg_ctda_fermi}}
\end{figure}

\begin{figure}[!!h]
\begin{center}
\includegraphics[width=7.90cm,angle=0]{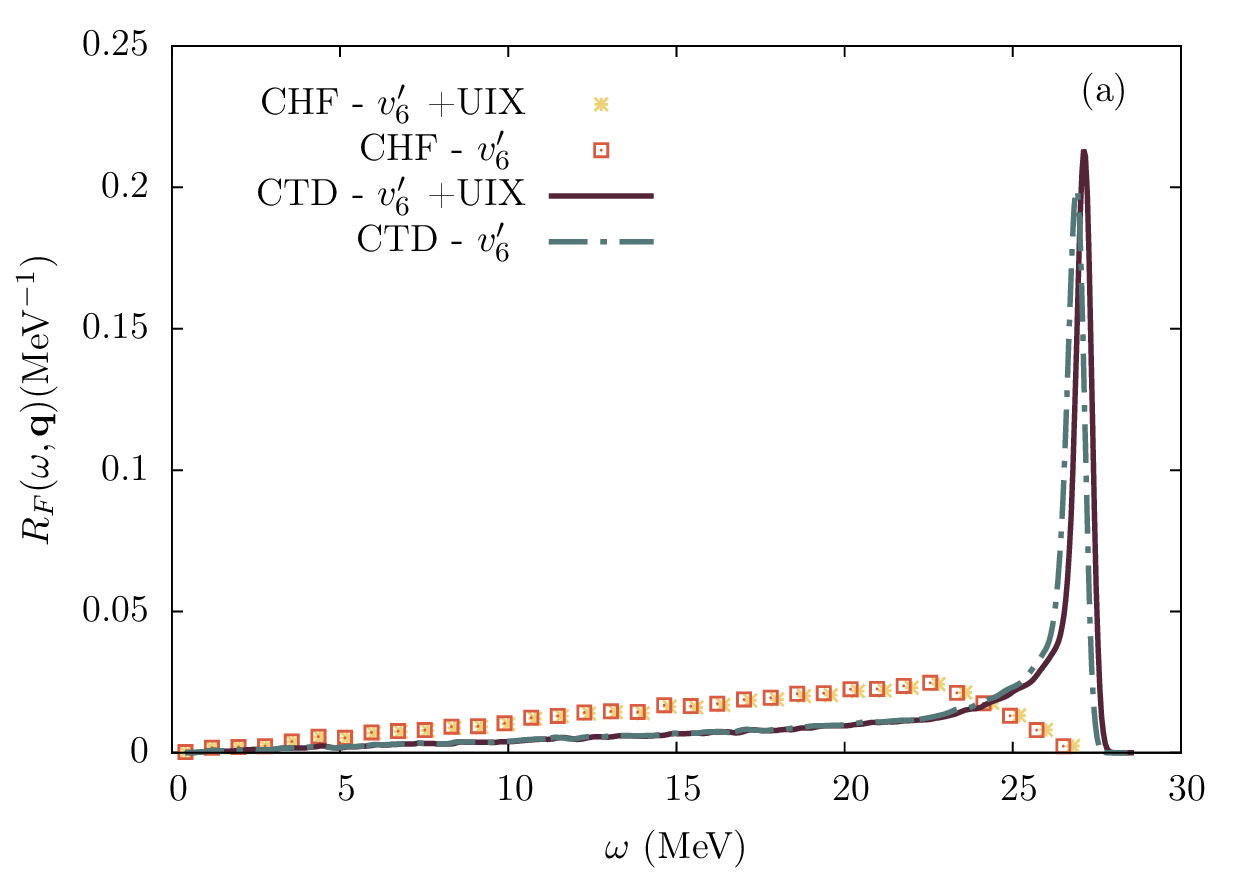}
\includegraphics[width=7.90cm,angle=0]{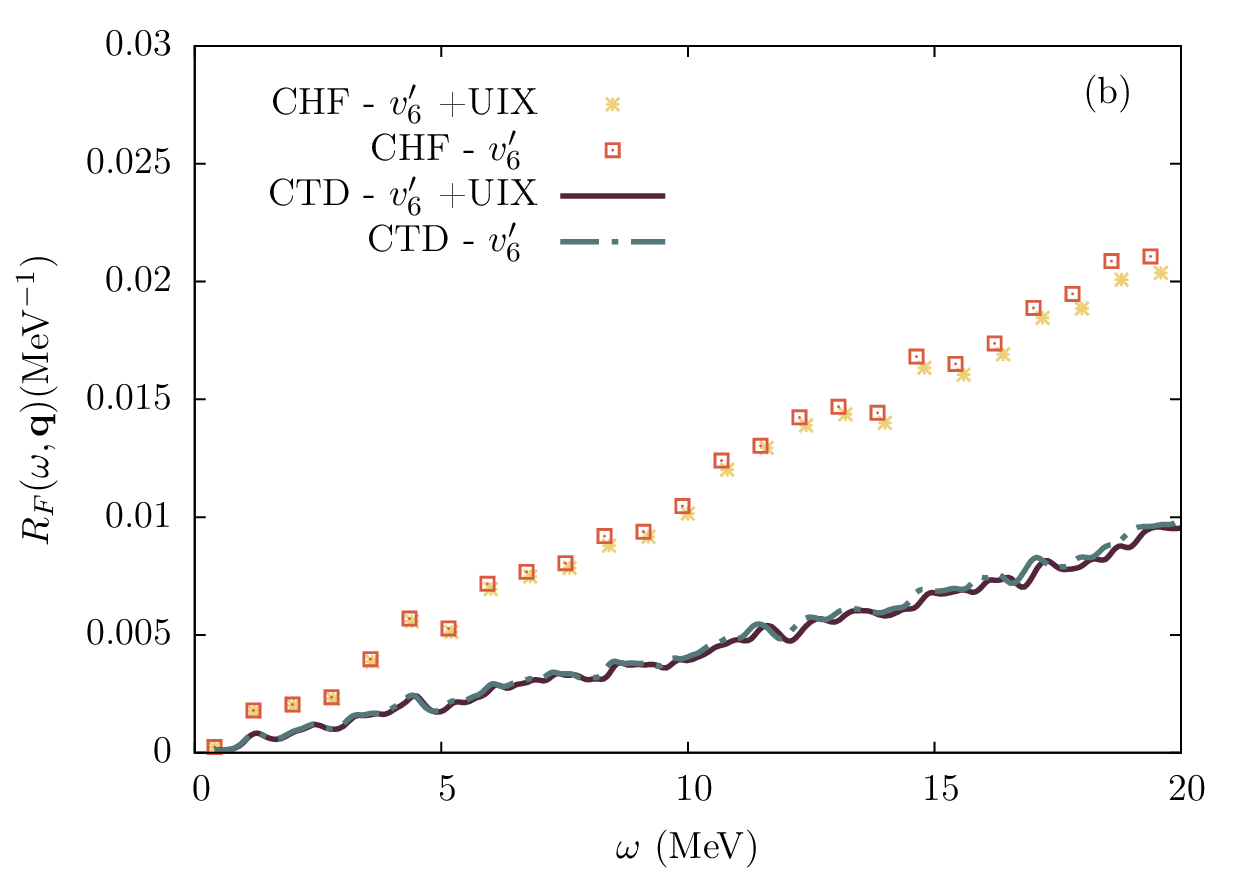}
\end{center}
\caption{The same as Fig. \ref{fig:cfg_ctda_fermi} but for Gamow-Teller response functions.\label{fig:cfg_ctda_gt}}
\end{figure}

The nuclear matter response calculated in CTD and CHF approximations for $|\mathbf{q}|=0.30$ MeV is displayed in Figs. \ref{fig:cfg_ctda_fermi} and \ref{fig:cfg_ctda_gt} for Fermi and Gamow-Teller transitions, respectively. By comparing with the findings of Refs. \cite{cowell_04,benhar_09}, it can be noticed that the peak corresponding to the collective mode is shifted to lower energies when the tree-body cluster is included. This effect, due to the change of the single particle energies, is only slightly mitigated when the UIX potential is included in the hamiltonian. 

Moreover, the three-body cluster produces a small depletion of the Fermi resonance at  $|\mathbf{q}|=0.30$, more apparent when the nuclear hamiltonian is lacking of the three-body potential. 

It is worth remarking that the two-body cluster results of Ref. \cite{cowell_04} have been obtained using the first six operators of the Argonne $v_{8}^\prime$ potential rather than the Argonne $v_{6}^\prime$ used in the present work. However, we have performed two-body cluster calculations with Argonne $v_{6}^\prime$ and the results are in extremely close to those of the truncated version of $v_{8}^\prime$.

\begin{figure}[!!h]
\begin{center}
\includegraphics[width=7.9cm,angle=0]{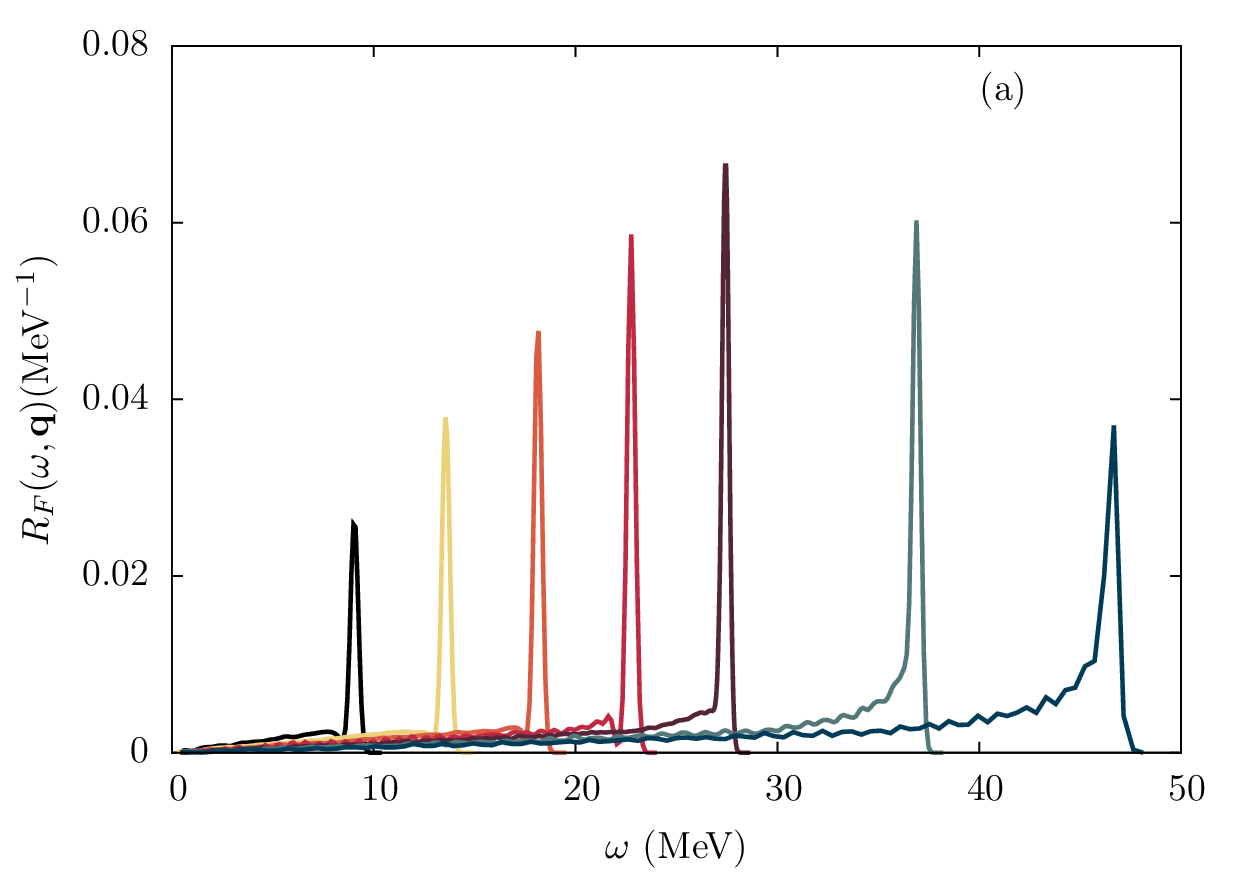}
\includegraphics[width=7.9cm,angle=0]{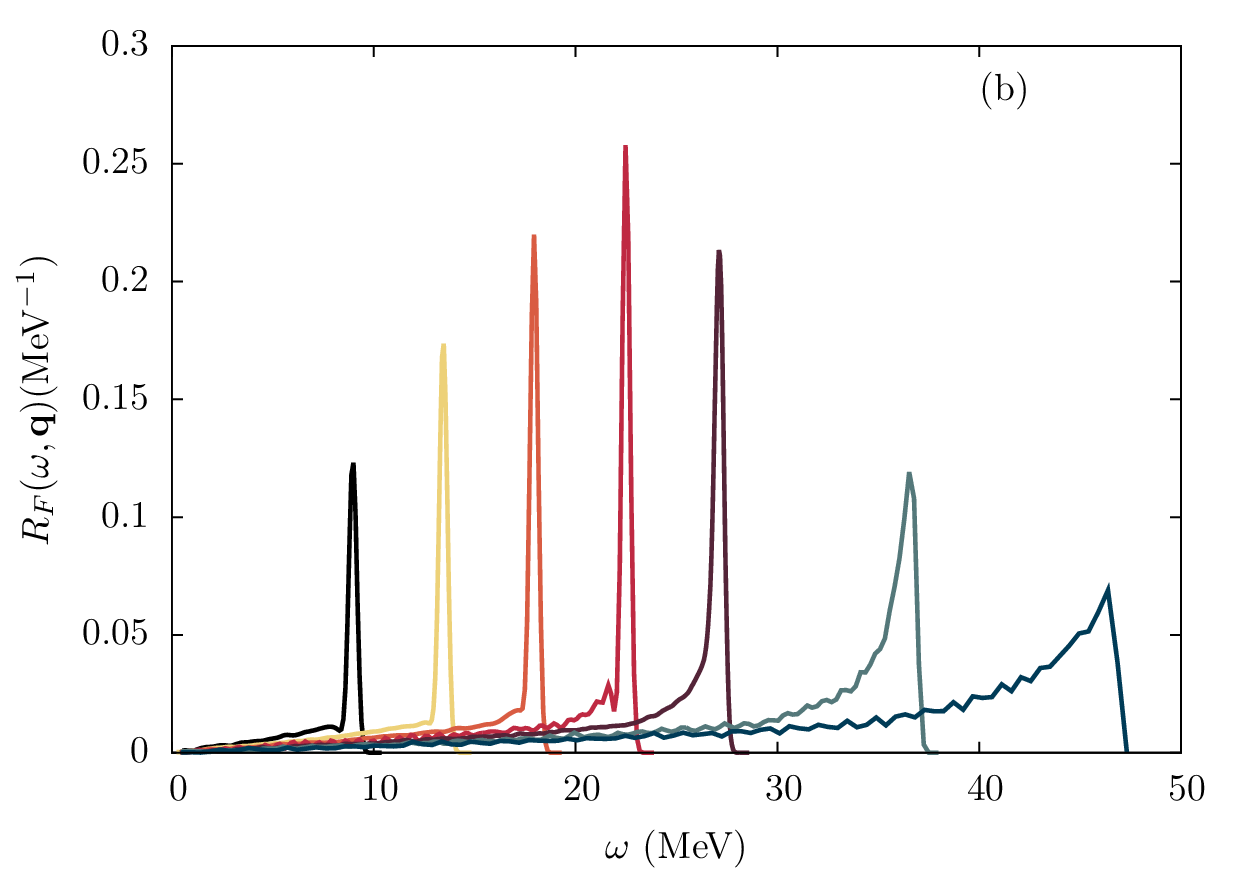}
\end{center}
\caption{Fermi (upper panel) Gamow-Teller (lower panel) response functions calculated at $q=0.10,0.15,0.20,0.25,0.30,0.40,0.50$ fm$^{-1}$ using $v_{6'}+$UIX  potential and correlations. \label{fig:q_scan_fgt}} 
\end{figure}

In Fig. \ref{fig:q_scan_fgt}, where the Fermi and Gamow-Teller responses are plotted for different values of  $|\mathbf{q}|$ ranging from $0.10$ fm$^{-1}$ to $0.50$ fm$^{-1}$.  The position of the highest peak of the Fermi responses shifts from $|\mathbf{q}|\simeq0.40$ fm$^{-1}$ of the two-body cluster approximation to $|\mathbf{q}|\simeq0.30$ fm$^{-1}$ of the present calculation. For the Gamow-Teller case, the three-body cluster contribution does not produce an appreciable shift, being the highest resonance remains peaked around $|\mathbf{q}|\simeq0.25$ fm$^{-1}$.

\subsection{Sum rules}
\label{sec:sum_rule}

The set of final states in Eq. (\ref{eq:resp_def}) is not exhausted by $1p-1h$ excitations. In principle, transitions to more complex multi $p-h$ states should also be considered. So far, the contribution of these states have been neglected, however, an estimate of their importance can be obtained computing the sum rules. 
The static structure function is defined by 
\begin{align}
S(\mathbf{q})&=\int d\omega S(\mathbf{q},\omega)\nonumber\\
&=\frac{1}{A}\int d\omega \sum_f |\langle \Psi_f | \hat{O}_{\textbf{q}}| \Psi_0\rangle |^2 \delta(\omega+E_0-E_n)\nonumber\\
&=\frac{1}{A} \langle \Psi_0 | \hat{O}_{\textbf{q}}^\dagger  \hat{O}_{\textbf{q}}| \Psi_0\rangle\, .
\end{align}
While a direct integration of the CTD response functions allows for the evaluation of $S(\mathbf{q})$, from the last line of the latter equation it is clear that the static structure function can be also evaluated by computing the ground state expectation value of $ \hat{O}_{\textbf{q}}^\dagger  \hat{O}_{\textbf{q}}$. Within the correlated basis function approach, the variational ground-state (VGS) expectation can be expressed in terms of the two-body operatorial distribution functions, calculated by means of the FHNC/SOC scheme in Ref. \cite{pandha_94}.

\begin{figure}[!b]
\begin{center}
\includegraphics[width=7.9cm,angle=0]{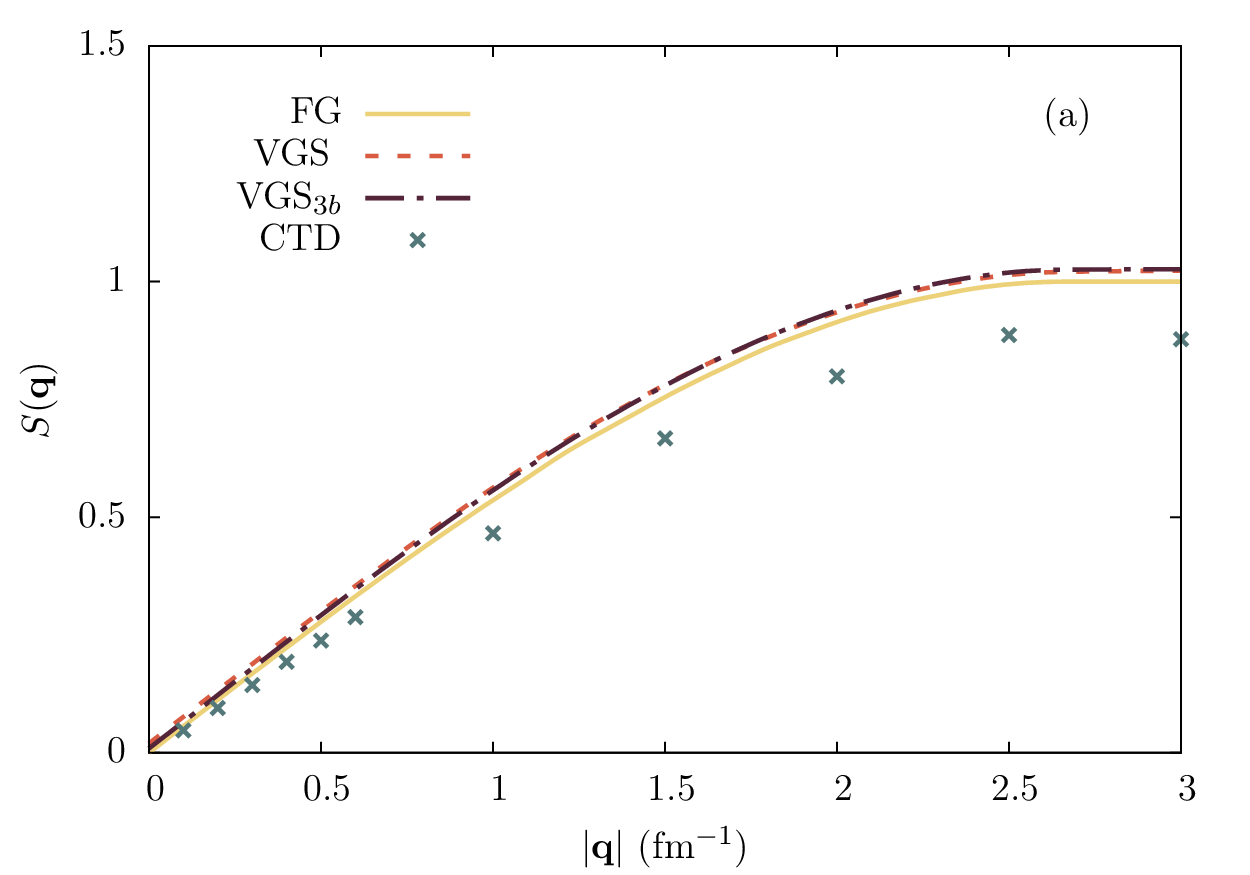}
\includegraphics[width=7.9cm,angle=0]{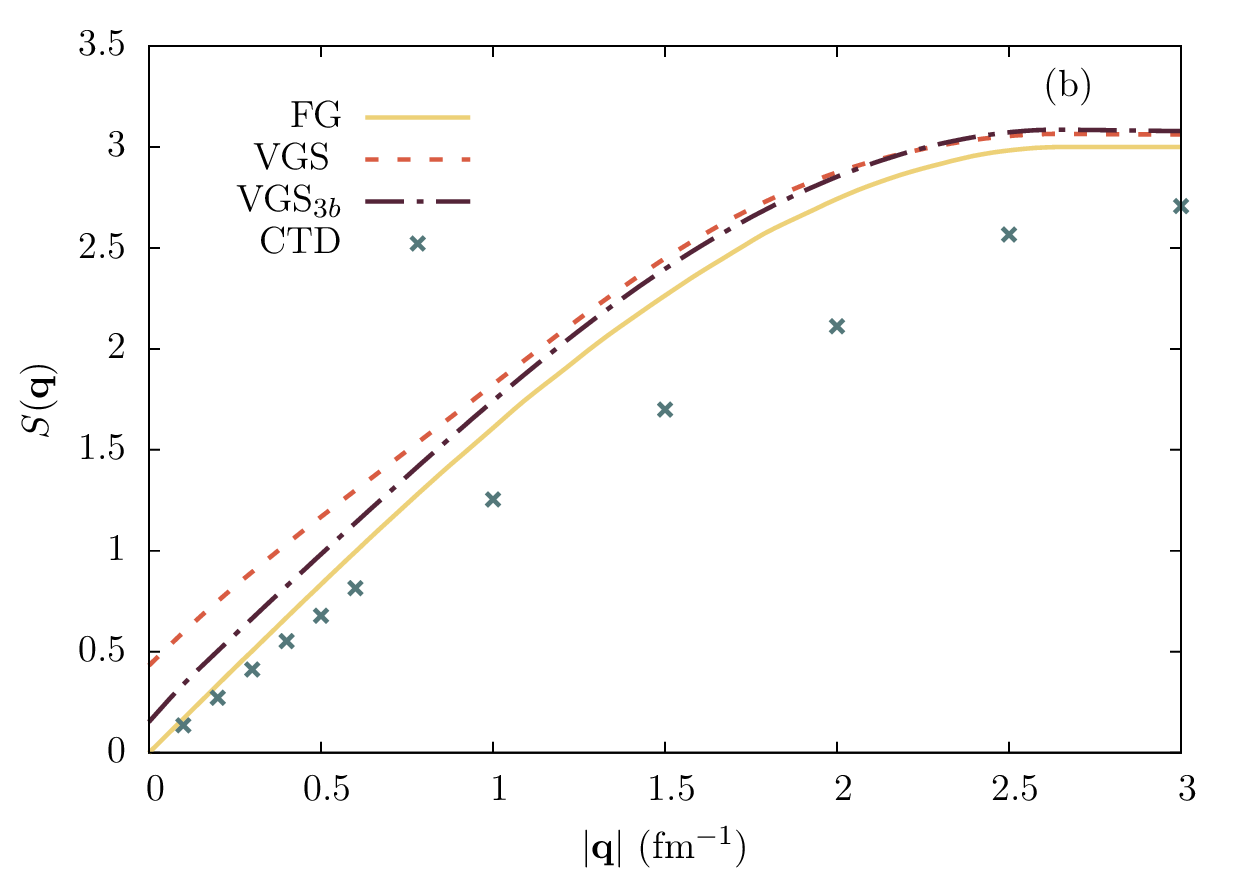}
\end{center}
\caption{Static response function for Fermi, panel (a), and Gamow-Teller, panel (b) transitions. Results are shown for a noninteracting FG (solid lines), for the integral of the CTD approximation of $S(\mathbf{q},\omega)$ (crosses) for the full FHNC/SOC VGS calculations (dashed lines) and for the three-body cluster approximation of the VGS (dotted-dashed lines). \label{fig:sumtot}}
\end{figure}

While the VGS calculations include all the multi $p-h$ excitations in the CBF basis, only the correlated $1p-1h$ states are taken into account in the CTD approximation. Therefore, multi $p-h$ contributions can be estimated from the difference $S^{VGS}(q)-S^{CTD}(q)$. 

Note that an interplay between many-body correlations and multi $p-h$ excitation could in principle take place. As a matter of fact, while VGS includes many-body correlations through the chain summations, in the CTDA of Ref. \cite{cowell_04} only two-body cluster terms have been considered. In the present work, three body cluster contributions have been taken into account in CTD calculations. Moreover, for the sake of isolating the effect of the multi $p-h$ excitations from the one of the many-body cluster contributions, we have also calculated the two-body operatorial distribution functions entering the variational ground state estimate of $S(\mathbf{q})$ at three-body cluster level.

The static structure functions for the Fermi and Gamow-Teller transitions are displayed in panel (a) and (b) of Fig. \ref{fig:sumtot}, respectively. Besides the FG curve, all the results have been obtained with an effective interaction that incorporates the the Argonne $v_{6}^\prime$+UIX potential and the corresponding correlations are used in the calculation of the effective weak-transition operators.

The curves corresponding to the Fermi transition are normalized in order for the sum rule of the non interacting FG to approach unity for large values of the momentum transfer. On the other hand, the Gamow-Teller results are normalized in such a way that both the transverse and longitudinal sum rules, to be defined in the following, tend to one for large $|\mathbf{q}|$.

As noted in Ref. \cite{cowell_04}, the VGS results obtained by computing the two-body distribution functions with FHNC/SOC summation scheme, because of the approximations involved in the calculation, do not fulfill the condition $S(0)=0$, required by baryon number conservation. The three-body cluster variational results, denoted by VGS$_{3b}$, also does not fulfill the baryon number conservation; however the reason for this most probably lies in the three-body cluster approximation which is known not to fulfill the sum rules of the two-body distribution functions. Conversely, the static structure function obtained within CTD approximation does exhibit the appropriate low-momentum limit. 

As far as the multi $p-h$ excitations are concerned, the two-body cluster results of Ref. \cite{cowell_04} show that their contribution is smaller than the dominant $1p-1h$ excitation, but it is not negligible. When three-body cluster is accounted for, the VGS and the CTD curves get closer. The shift turns out to be detectable for the Fermi transition case, while for the total Gamow-Teller response is very small. Considering also the fact that $VGS_{3b}$ results lie well above the CTD crosses, we may conclude that the difference between variational ground state and correlated Tamm-Dancoff results has largely to be ascribed to the multi $p-h$ excitations. 

\begin{figure}[!!h]
\begin{center}
\includegraphics[width=8cm,angle=0]{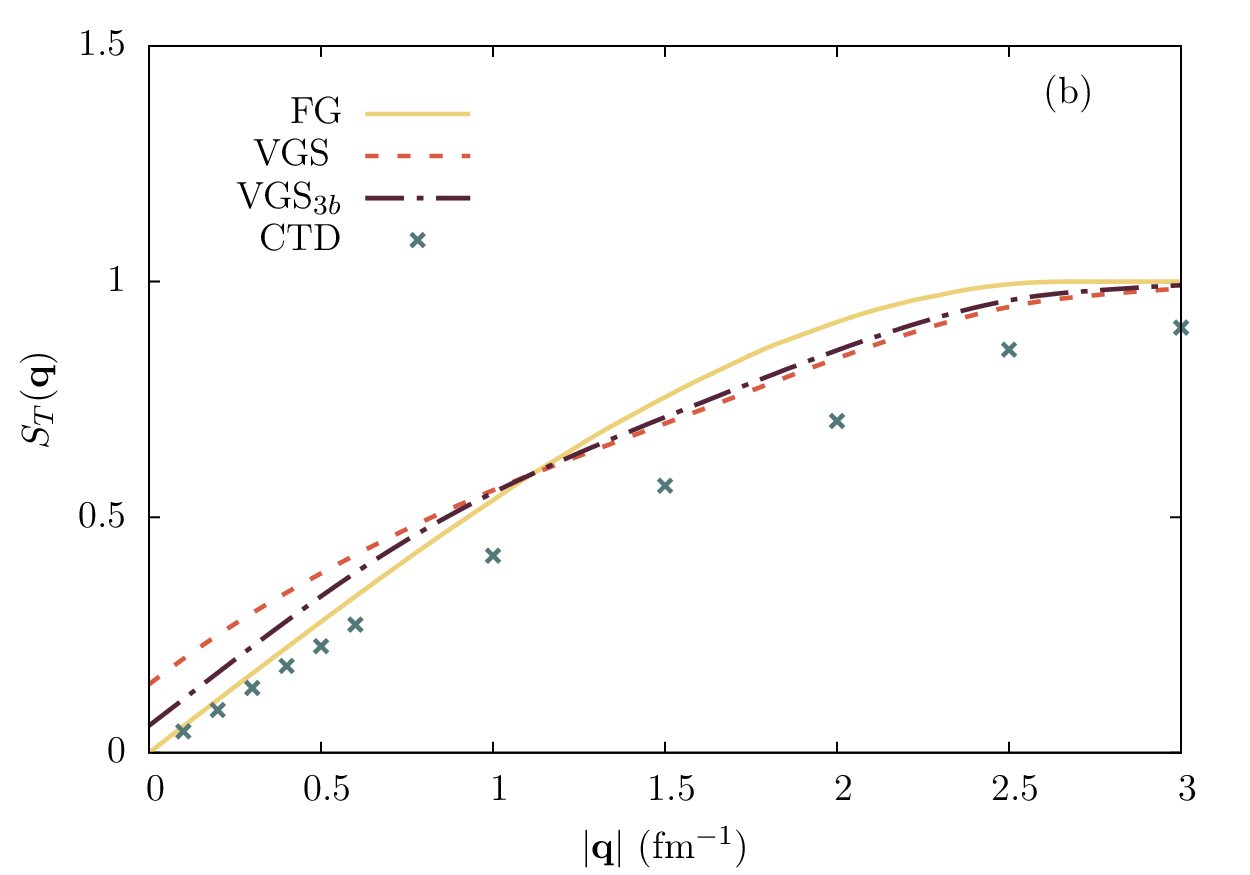}
\includegraphics[width=8cm,angle=0]{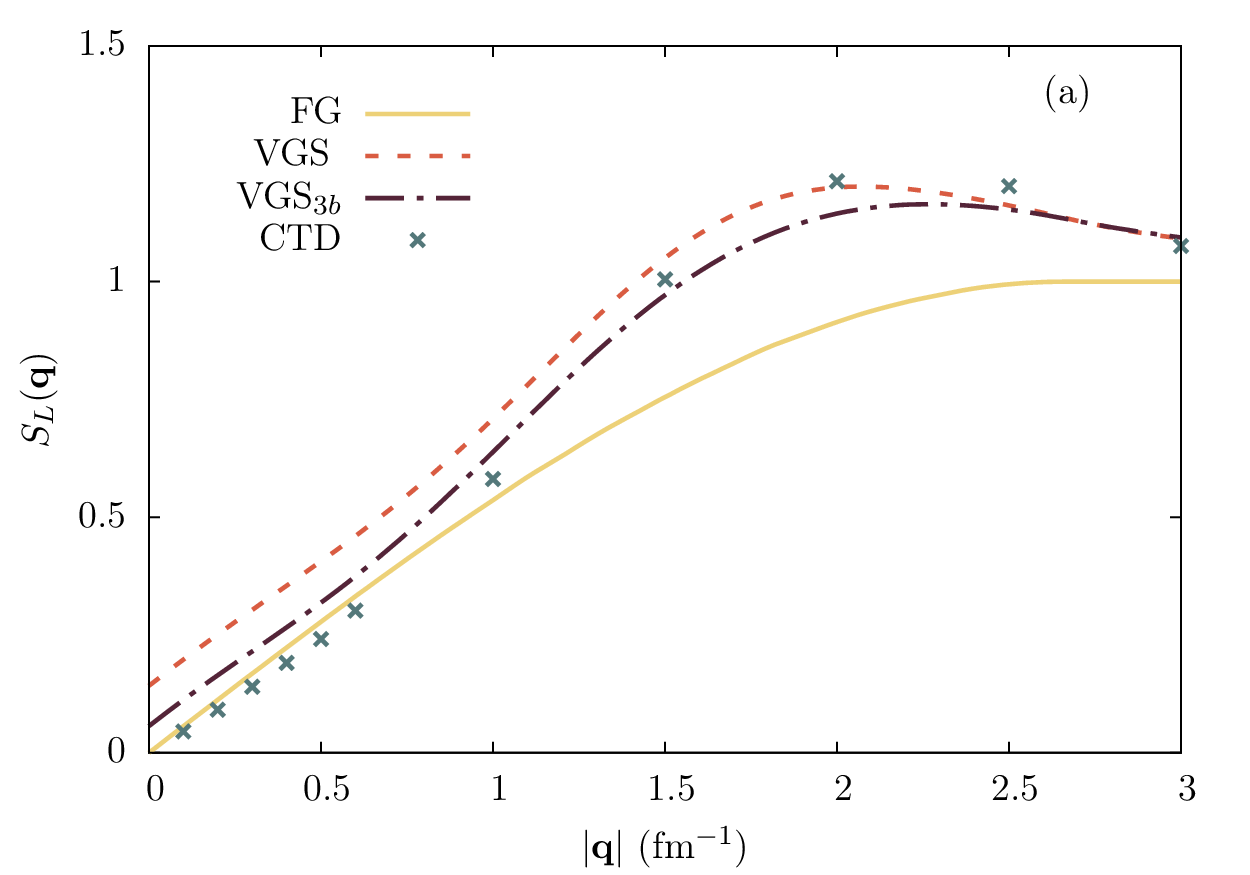}
\end{center}
\caption{Spin longitudinal (left panel) and spin transverse (right panel) static response functions, calculated using VGS (solid lines), two-body CTDA (squares), two-body CTDA (stars), and noninteracting FG (dotted lines). \label{fig:sum_lt}}
\end{figure}

There are experimental and theoretical indications that the spin longitudinal and spin transverse response functions can differ significantly due to tensor forces. Thus, we studied how the UIX three-body force affects these quantities computing the spin transverse and spin longitudinal static structure functions, defined as
\begin{align}
S^T(q)&=\frac{1}{A}\int d\omega \sum_n |\langle \Psi_n | \hat{q} \wedge \hat{O}_{\textbf{q}}^{GT}| \Psi_0\rangle |^2 \delta(\omega+E_0-E_n)\nonumber\\
S^L(q)&=\frac{1}{A}\int d\omega \sum_n |\langle \Psi_n | \hat{q} \cdot \hat{O}_{\textbf{q}}^{GT}| \Psi_0\rangle |^2 \delta(\omega+E_0-E_n)
\end{align}
A comparison between Fig. \ref{fig:sum_lt} and Fig. 17 of Ref. \cite{cowell_04} shows that the inclusion of the UIX potential brings the CTD curves for $S^T(q)$ closer to those of VGS across all the values of $|\mathbf{q}|$. For relatively large momentum transfer it is clear that the difference between the variational ground-state results and the CTD calculations is due to $2p-2h$ excitations.  The curves VGS and VGS$_{3b}$ are almost superimposed in this region, indicating that the contribution of many-body clusters is very small.

As far as the longitudinal static response function is concerned, the position of the maximum of the CTD calculations including UIX potential almost coincide with the VGS and VGS$_{3b}$ results. At small momentum transfer however, the three-body cluster CTD points lie below the variational results, mostly because of deficiencies of the latter, unable to fulfill the sum rules. 

Recently \cite{shen_12}, the sum rules relations (additional sum rules with an increasing power of $\omega$ in the integrand can be defined) has been inverted to obtain the spin-response function of PNM at zero momentum transfer. The authors of Ref. \cite{shen_12} used the formalism of AFDMC to compute the operatorial distribution functions, obtaining very promising results. As a follow up of the present work, we are planning to compute the response function of PNM and compare it with their findings. 

\section{Conclusions}
\label{sec:concl}
We have shown that the weak response of isospin symmetric nuclear matter, obtained from the effective operators computed in the two-body cluster approximation, exhibits a 
sizable dependence on the choice of the correlation function, which is in fact unphysical. 
When three-body clusters diagrams are considered, the transition matrix elements, in which correlations enter only through the effective weak operators, become 
nearly independent of the correlation functions. 

The three-body cluster contributions, described in Section \ref{sec:ei}, have been consistently included in the construction of the effective interaction. As a first step we have computed the EoS of SNM for an hamiltonian including the two-body potential only. In this case, the three-body cluster effective interaction provides a EoS of SNM much closer to to the one 
resulting from full FHNC/SOC calculations, compared to the one obtained using the two-body cluster effective interaction of Ref. \cite{cowell_04, benhar_07}.
The leading contributions of the UIX potential, emerging at three-body cluster level, have been also included in the effective potential. As a result 
the EoS of SNM exhibits saturation at $\rho\simeq 0.18\,\text{fm}^{-1}$.

Inclusion of the three-nucleon interactions also affects the single particle spectrum, leading in turn to a shift of the CHF response as a function of energy transfer.

The main effect of the three-nucleon force on the TDA response originates from a change of the off-diagonal elements of the effective interaction. As a result, the collective mode associated with the 
Fermi transition at $|\mathbf{q}|=0.3$ MeV turns out to be shifted to lower energy, although its magnitude is nearly unaffected by the three-body cluster contributions. On the other hand, a depletion of the peak is observed for the Gamow-Teller transition. The analysis of the TDA response for different momentum transfer also reveals a sizable effect of three-nucleon cluster
contributions.

The sum rules for the Fermi transition comes closer to the variational results once the three-body cluster cluster is taken into account, thus confirming the importance of many body effects, 
which are included in the variational calculations through the chain summations. The residual discrepancy cannot be accounted for by the $n>4-$body cluster contributions, and is 
likely to be ascribable to the effect of multi $p-h$ excitations, which are taken into account in variational calculations.
This effect appears to be even larger in the structure function obtained from the Gamow-Teller response. In this case the change due to the three-body cluster is in fact very small. 

Some improvements are observed in the sum rules of the longitudinal and the transverse response. As shown in Fig. \ref{fig:sum_lt},  the results of the three-body cluster calculations of $S^L(q)$ carried out within TDA are slightly closer to the variational ones for all the values of $|\mathbf{q}|$,  compared to the two-body cluster case. Moreover, the position of the maximum of the TDA calculations of the static transverse response is almost coincident with that obtained from variational calculations. At small momentum transfer however, the three-body cluster contributions move
the TDA $S^T(q)$ away from the variational result.


\begin{thebibliography}{25}
\expandafter\ifx\csname natexlab\endcsname\relax\def\natexlab#1{#1}\fi
\providecommand{\url}[1]{\texttt{#1}}
\providecommand{\href}[2]{#2}
\providecommand{\path}[1]{#1}
\providecommand{\eprint}[1]{\href{http://arxiv.org/abs/#1}{\path{#1}}}
\providecommand{\DOIprefix}{doi:}
\providecommand{\ArXivprefix}{arXiv:}
\providecommand{\urlprefix}{URL: }
\providecommand{\Pubmedprefix}{pmid:}
\providecommand{\DOI}[1]{\href{http://dx.doi.org/#1}{\path{#1}}}
\providecommand{\Pubmed}[1]{\href{pmid:#1}{\path{#1}}}
\providecommand{\bibinfo}[2]{#2}
\ifx\xfnm\relax \def\xfnm[#1]{\unskip,\space#1}\fi
\bibitem[{Cowell and Pandharipande(2004)}]{cowell_04}
\bibinfo{author}{S.~Cowell}, \bibinfo{author}{V.~R. Pandharipande},
  \bibinfo{journal}{Phys. Rev. C} \bibinfo{volume}{70} (\bibinfo{year}{2004})
  \bibinfo{pages}{035801}. 
\bibitem[{Benhar and Farina(2009)}]{benhar_09}
\bibinfo{author}{O.~Benhar}, \bibinfo{author}{N.~Farina},
  \bibinfo{journal}{Physics Letters B} \bibinfo{volume}{680}
  (\bibinfo{year}{2009}) \bibinfo{pages}{305 -- 309}. 
\bibitem[{Fantoni and Pandharipande(1987)}]{fantoni_87}
\bibinfo{author}{S.~Fantoni}, \bibinfo{author}{V.~Pandharipande},
  \bibinfo{journal}{Nuclear Physics A} \bibinfo{volume}{473}
  (\bibinfo{year}{1987}) \bibinfo{pages}{234 -- 266}. 
\bibitem[{Fabrocini and Fantoni(1989)}]{fabrocini_89}
\bibinfo{author}{A.~Fabrocini}, \bibinfo{author}{S.~Fantoni},
  \bibinfo{journal}{Nuclear Physics A} \bibinfo{volume}{503}
  (\bibinfo{year}{1989}) \bibinfo{pages}{375 -- 403}. 
\bibitem[{Fabrocini(1997)}]{fabrocini_97}
\bibinfo{author}{A.~Fabrocini}, \bibinfo{journal}{Phys. Rev. C}
  \bibinfo{volume}{55} (\bibinfo{year}{1997}) \bibinfo{pages}{338--348}.
\bibitem[{Wiringa et~al.(1995)Wiringa, Stoks, and Schiavilla}]{wiringa_95}
\bibinfo{author}{R.~B. Wiringa}, \bibinfo{author}{V.~G.~J. Stoks},
  \bibinfo{author}{R.~Schiavilla}, \bibinfo{journal}{Phys. Rev. C}
  \bibinfo{volume}{51} (\bibinfo{year}{1995}) \bibinfo{pages}{38--51}.
\bibitem[{Lagaris and Pandharipande(1981)}]{lagaris_81}
\bibinfo{author}{I.~Lagaris}, \bibinfo{author}{V.~Pandharipande},
  \bibinfo{journal}{Nuclear Physics A} \bibinfo{volume}{359}
  (\bibinfo{year}{1981}) \bibinfo{pages}{349 -- 364}. 
\bibitem[{Pudliner et~al.(1995)Pudliner, Pandharipande, Carlson, and
  Wiringa}]{pudliner_95}
\bibinfo{author}{B.~S. Pudliner}, \bibinfo{author}{V.~R. Pandharipande},
  \bibinfo{author}{J.~Carlson}, \bibinfo{author}{R.~B. Wiringa},
  \bibinfo{journal}{Phys. Rev. Lett.} \bibinfo{volume}{74}
  (\bibinfo{year}{1995}) \bibinfo{pages}{4396--4399}. 
\bibitem[{Clark and Westhaus(1966)}]{clark_66}
\bibinfo{author}{J.~W. Clark}, \bibinfo{author}{P.~Westhaus},
  \bibinfo{journal}{Phys. Rev.} \bibinfo{volume}{141} (\bibinfo{year}{1966})
  \bibinfo{pages}{833--857}. 
\bibitem[{Fantoni and Fabrocini(1998)}]{fantoni_98}
\bibinfo{author}{S.~Fantoni}, \bibinfo{author}{A.~Fabrocini}, in:
  \bibinfo{editor}{J.~Navarro}, \bibinfo{editor}{A.~Polls} (Eds.),
  \bibinfo{booktitle}{Microscopic Quantum Many-Body Theories and Their
  Applications}, volume \bibinfo{volume}{510} of
  \textit{\bibinfo{series}{Lecture Notes in Physics}},
  \bibinfo{publisher}{Springer Berlin / Heidelberg}, \bibinfo{year}{1998}, pp.
  \bibinfo{pages}{119--186}. 
  \bibinfo{note}{10.1007/BFb0104526}.
\bibitem[{Pandharipande(1971)}]{pandharipande_71}
\bibinfo{author}{V.~Pandharipande}, \bibinfo{journal}{Nuclear Physics A}
  \bibinfo{volume}{174} (\bibinfo{year}{1971}) \bibinfo{pages}{641 -- 656}.
\bibitem[{Ristig et~al.(1971)Ristig, Louw, and Clark}]{ristig_71}
\bibinfo{author}{M.~L. Ristig}, \bibinfo{author}{W.~J.~T. Louw},
  \bibinfo{author}{J.~W. Clark}, \bibinfo{journal}{Phys. Rev. C}
  \bibinfo{volume}{3} (\bibinfo{year}{1971}) \bibinfo{pages}{1504--1513}.
\bibitem[{Pandharipande(1972)}]{pandharipande_72}
\bibinfo{author}{V.~Pandharipande}, \bibinfo{journal}{Nuclear Physics A}
  \bibinfo{volume}{181} (\bibinfo{year}{1972}) \bibinfo{pages}{33 -- 48}.
\bibitem[{Lovato et~al.(2011)Lovato, Benhar, Fantoni, Illarionov, and
  Schmidt}]{lovato_11}
\bibinfo{author}{A.~Lovato}, \bibinfo{author}{O.~Benhar},
  \bibinfo{author}{S.~Fantoni}, \bibinfo{author}{A.~Y. Illarionov},
  \bibinfo{author}{K.~E. Schmidt}, \bibinfo{journal}{Phys. Rev. C}
  \bibinfo{volume}{83} (\bibinfo{year}{2011}) \bibinfo{pages}{054003}.
\bibitem[{Cowell and Pandharipande(2003)}]{cowell_03}
\bibinfo{author}{S.~Cowell}, \bibinfo{author}{V.~R. Pandharipande},
  \bibinfo{journal}{Phys. Rev. C} \bibinfo{volume}{67} (\bibinfo{year}{2003})
  \bibinfo{pages}{035504}. 
\bibitem[{Benhar and Valli(2007)}]{benhar_07}
\bibinfo{author}{O.~Benhar}, \bibinfo{author}{M.~Valli},
  \bibinfo{journal}{Phys. Rev. Lett.} \bibinfo{volume}{99}
  (\bibinfo{year}{2007}) \bibinfo{pages}{232501}. 
\bibitem[{Lagaris and Pandharipande(1980)}]{lagaris_80}
\bibinfo{author}{I.~Lagaris}, \bibinfo{author}{V.~Pandharipande},
  \bibinfo{journal}{Nuclear Physics A} \bibinfo{volume}{334}
  (\bibinfo{year}{1980}) \bibinfo{pages}{217 -- 228}. 
\bibitem[{Morales et~al.(2002)Morales, Pandharipande, and
  Ravenhall}]{morales_02}
\bibinfo{author}{J.~Morales}, \bibinfo{author}{V.~R. Pandharipande},
  \bibinfo{author}{D.~G. Ravenhall}, \bibinfo{journal}{Phys. Rev. C}
  \bibinfo{volume}{66} (\bibinfo{year}{2002}) \bibinfo{pages}{054308}.
\bibitem[{Lovato(2012)}]{lovato_12c}
\bibinfo{author}{A.~Lovato}, \bibinfo{title}{Ab initio calculations on nuclear
  matter properties including the effects of three-nucleons interaction}, Ph.D.
  thesis, SISSA-ISAS Trieste, \bibinfo{year}{2012}.
\bibitem[{Farina(2009)}]{farina_09}
\bibinfo{author}{N.~Farina}, \bibinfo{title}{Weak Response of Nuclear Matter},
  Ph.D. thesis, Sapienza Universit\`a di Roma, \bibinfo{year}{2009}.
  \ArXivprefix  \eprint{0901.2507}.
\bibitem[{Jastrow(1955)}]{jastrow_55}
\bibinfo{author}{R.~Jastrow}, \bibinfo{journal}{Phys.Rev.} \bibinfo{volume}{98}
  (\bibinfo{year}{1955}) \bibinfo{pages}{1479--1484}.
\bibitem[{Fetter and Walecka(2003)}]{fetter_03}
\bibinfo{author}{A.~Fetter}, \bibinfo{author}{J.~Walecka},
  \bibinfo{title}{Quantum Theory of Many-Particle Systems}, Dover Books on
  Physics, \bibinfo{publisher}{Dover Publications}, \bibinfo{year}{2003}.
\bibitem[{Akmal(1998)}]{akmal_98}
\bibinfo{author}{A.~Akmal}, \bibinfo{title}{Variational studies of nucleon
  matter with realistic potentials}, Ph.D. thesis, University of Illinois at
  {U}rbana-Champaign, \bibinfo{year}{1998}.
\bibitem[{Pandharipande et~al.(1994)Pandharipande, Carlson, Pieper, Wiringa,
  and Schiavilla}]{pandha_94}
\bibinfo{author}{V.~R. Pandharipande}, \bibinfo{author}{J.~Carlson},
  \bibinfo{author}{S.~C. Pieper}, \bibinfo{author}{R.~B. Wiringa},
  \bibinfo{author}{R.~Schiavilla}, \bibinfo{journal}{Phys. Rev. C}
  \bibinfo{volume}{49} (\bibinfo{year}{1994}) \bibinfo{pages}{789--801}.
\bibitem[{Shen et~al.(2012)Shen, Gandolfi, Reddy, and Carlson}]{shen_12}
\bibinfo{author}{G.~Shen}, \bibinfo{author}{S.~Gandolfi},
  \bibinfo{author}{S.~Reddy}, \bibinfo{author}{J.~Carlson}
  (\bibinfo{year}{2012}). \ArXivprefix  \eprint{1205.6499}.

\end{thebibliography}

\end{document}